\documentclass[a4paper,12pt]{article}

\usepackage{graphicx}
\usepackage{amsmath}
\usepackage{amssymb}
\usepackage{hyperref}
\usepackage{tabularx}
\newcolumntype{M}{>{$}c<{$}}
\usepackage[matrix,arrow,curve]{xy}

\numberwithin{equation}{section}
\numberwithin{figure}{section}
\numberwithin{table}{section}

\def\papertitlepage{\baselineskip 3.5ex\thispagestyle{empty}}
\def\Title#1{\baselineskip 1cm \vspace{1.5cm}%
  \begin{center}{\Large\bf #1}\end{center}\vspace{0.5cm}}
\def\Authors#1{\begin{center}\renewcommand{\thefootnote}{\fnsymbol{footnote}}{\it #1}\end{center}}
\def\Abstract{\vspace{1.0cm}%
  \begin{center}{\large\bf Abstract}\end{center}}

\renewenvironment{thebibliography}{\pagebreak[3]\par\vspace{0.6em}
\begin{flushleft}{\large \bf References}\end{flushleft}
\vspace{-1.0em}

\begin{enumerate}\if@twocolumn\baselineskip=0.6em\itemsep -0.2em
\else\itemsep -0.2em\fi\labelsep 0.1em}{\end{enumerate} }

\def\calA{{\cal A}} \def\calB{{\cal B}} 
  \def\calF{{\cal F}}
 \def\calH{{\cal H}}

 \def\calZ{{\cal Z}}

\def\del{\partial}

\def\mathe{\mathrm{e}}
\def\mathi{i}

\def\Tr{\mathrm{Tr}}
\def\tr{\mathrm{tr}}

\DeclareMathDelimiter{\lcolon}{\mathopen}{operators}{"3A}{largesymbols}{"3A}
\DeclareMathDelimiter{\rcolon}{\mathclose}{operators}{"3A}{largesymbols}{"3A}
\def\NO#1{\lcolon#1\rcolon}

\def\+{\!\!+\!\!}
\def\delcech{\Check{\delta}}
\def\deldol{\overline{\del}_{X}}
\def\targetd{\mathrm{d}}
\def\targetdbar{\overline{\del}}

\def\lpar{(\!(}
\def\rpar{)\!)}
\def\lbra{[\![}
\def\rbra{]\!]}

\def\dynkin(#1){(#1)}

\def\bra<#1|{\langle#1|}
\def\ket|#1>{|#1\rangle}
\def\braket<#1|#2>{\langle#1|#2\rangle}
\def\llangle{\langle\!\langle}
\def\rrangle{\rangle\!\rangle}
\def\bbra<#1|{\llangle#1|}
\def\kket|#1>{|#1\rrangle}
\def\bbraket<#1|#2>{\llangle#1|#2\rrangle}
\def\conj{\ast}

\begin{document}
{\papertitlepage
\vspace*{0cm}
{\hfill
\begin{minipage}{4.2cm}
IFT-P.009/2008\par\noindent June, 2008
\end{minipage}}
\Title{Hilbert space of curved $\beta\gamma$ systems \\ on quadric cones}
\Authors{{\sc Yuri Aisaka\footnote{\tt yuri@ift.unesp.br}}
and
{\sc E.~Aldo~Arroyo\footnote{\tt aldohep@ift.unesp.br}}
\\
Instituto de F\'{i}sica Te\'{o}rica, State University of S\~{a}o Paulo, \\[-2ex]
Rua Pamplona 145, S\~{a}o Paulo, SP 01405-900, Brasil
}

} 

\vskip-\baselineskip
{\baselineskip .5cm \Abstract
We clarify the structure of the
Hilbert space of curved $\beta\gamma$ systems defined by a
quadratic constraint. The constraint is studied using
intrinsic and BRST methods, and their partition functions
are shown to agree. 
The quantum BRST cohomology is non-empty only at ghost numbers 0 and 1,
and there is a one-to-one mapping between these two sectors.
In the intrinsic description, the ghost number 1
operators correspond to the ones that are not globally defined on
the constrained surface. Extension of the results to the pure
spinor superstring is discussed in a separate work. }
\newpage
\setcounter{footnote}{0}

\tableofcontents

\section{Introduction}

About seven years ago, a new formalism for the superstring which
achieves manifest ten dimensional super-Poincar\'{e} covariance was
proposed~\cite{Berkovits:2000fe}.
As of today, the formalism has passed various consistency checks and
has been used to compute multiloop amplitudes and to describe
Ramond-Ramond backgrounds in a super-Poincar\'{e} covariant manner.

One of the key ingredients of the formalism is the use of a bosonic
variable $\lambda^{\alpha}$ that is constrained non-linearly to be a pure spinor
$\lambda\gamma^{\mu}\lambda=0$.
In a sense, $\lambda^{\alpha}$
can be thought as the ghost for the Green-Schwarz-Siegel worldsheet constraint $d_{\alpha}$.
Although the use of such a constrained ghost system is unconventional,
it can be used to construct vertex operators
and to define string amplitudes as worldsheet correlation functions%
~\cite{Berkovits:2000fe,Berkovits:2004px,Berkovits:2006vi}.
Dependence of the amplitudes on the non-zero modes of $\lambda^{\alpha}$ and its conjugate $\omega_{\alpha}$
is fixed by the operator product expansions,
and the functional integral over the zero-modes can be inferred
by requiring BRST and super-Poincar\'{e} invariance.

Although the basic ingredients for computing on-shell amplitudes
are already there, it would be useful to understand
the functional integral over $\lambda^{\alpha}$ without relying on the BRST invariance,
or equivalently, to understand the nature of the Hilbert space
in the operator formalism.
This would be necessary, for example,
if one wishes to apply the formalism to construct a super string field theory.

There are two basic strategies to study the structure of
the Hilbert space for the pure spinors.
The first is to deal directly with the constrained variables,
and define the Hilbert space as the space of operators
that are consistent with the pure spinor constraint~\cite{Berkovits:2000fe}.
To be consistent with the constraint,
the conjugate $\omega_{\alpha}$ has to appear in combinations invariant under the
``gauge transformations'' $\delta_{\Lambda}\omega_{\alpha}=\Lambda^{\mu}(\gamma_{\mu}\lambda)_{\alpha}$
generated by the constraint $\lambda\gamma_{\mu}\lambda$.
The other is to try to remove the constraint by introducing BRST ghosts.
The constraint is then expressed effectively as the cohomology condition
of the BRST operator $D$~\cite{Chesterman:2004xt}\footnote{%
$D$ should not to be confused with the ``physical'' BRST operator $Q=\int \lambda^{\alpha}d_{\alpha}$
of the pure spinor formalism.
(Because a possible use of $D$ is to combine it with $Q$ to construct
a single nilpotent operator $\Hat{Q}=D+Q+\cdots$,
we called $D$ a ``mini-BRST'' operator in~\cite{PS}.)
}.

Each method has its own advantages and disadvantages at the present time.
For the first method, the theory of so-called curved $\beta\gamma$ systems provide
a natural framework to deal with
the constraint~\cite{Malikov:1998dw}\cite{Kapustin:2005pt,Witten:2005px,Nekrasov:2005wg}.
The basic idea is to regard the pure spinor sector as a collection of free bosonic $\beta\gamma$ systems
defined locally but intrinsically on the pure spinor target space.
Although this \v{C}ech type formulation provides a nice description of the pure spinor sector,
self-contained rules for performing the functional integral over the fields
defined only locally remains to be clarified.
The BRST method for the pure spinor system, on the other hand,
meets more severe difficulties.
Since the pure spinor constraint is infinitely reducible
(meaning there are relations among the constraints, and relations-for-relations and so on)
one has to introduce an infinite chain of ghosts-for-ghosts~\cite{Chesterman:2004xt}.
Although the infinite ghosts in fact are fairly useful for computing
partition functions~\cite{Berkovits:2005hy,PS},
expressions for the vertex operators and
the composite reparameterization $b$-ghost
become complicated and at best rather formal.

\bigskip
Taking aim at clarifying the Hilbert space for the pure spinors,
we in this paper consider models with a single irreducible
constraint $\lambda^{i}\lambda^{i}=0$ ($\lambda^{i}\not\equiv0$, $i=1\sim N$).
It will be argued that
the curved $\beta\gamma$ and BRST formalisms
provide equivalent classical descriptions of the system,
although, quantum mechanically,
the Hilbert spaces of the two descriptions
differ slightly due to the different normal ordering prescriptions used.
Nevertheless, since our partition function
(in fact an index $\Tr[(-1)^{F}\cdots]$)
is defined so that it is insensitive to quantum corrections,
the two descriptions lead to the same partition function
even quantum mechanically.
We shall use the partition function as a guide to study
the structure of the Hilbert space.

The BRST formalism is designed so that the
ghost number $0$ cohomology of the BRST operator
$D=\int b(\lambda\lambda)$ reproduces the usual gauge invariant operators,
that is,
the gauge invariant polynomials made out of $\lambda^{i}$ and its conjugate $\omega_{i}$,
and their derivatives.
In the curved $\beta\gamma$ language, those gauge invariant polynomials
are nothing but the globally defined operators\footnote{%
In lower dimensions $N\le3$, there are globally defined operators
which cannot be described as gauge invariant polynomials
\cite{Grassi:2007va}. }, so one expects the agreement on the ghost
number $0$ sector as has been noted in~\cite{Grassi:2006wh}. In
this paper, we will claim that the equivalence goes beyond the
ghost number $0$ sector. For example, the BRST ghost itself $b$
(ghost number $+1$) is clearly in the cohomology of $D=\int
b(\lambda\lambda)$. In the curved $\beta\gamma$ description, $b$
will be identified as an operator that is defined only on single
overlaps of the coordinate charts, or in other words, as an
element of the first \v{C}ech cohomology.

The fact that the number of coordinate overlap corresponds to the BRST ghost number
can be best understood in the so-called non-minimal or Dolbeault formulation
of the curved $\beta\gamma$ systems.
In this formulation, one introduces the complex conjugate $\overline{\lambda}_{i}$ of $\lambda^{i}$
and its differential $r_{i}=\targetd\overline{\lambda}_{i}$, together with their conjugates
$\overline{\omega}^{i}$ and $s^{i}$.
The relevant cohomology operator is an extension of the Dolbeault differential
$\deldol=-r_{i}\overline{\omega}^{i}\sim \targetd\overline{\lambda}_{i}(\del/\del\overline{\lambda}_{i})$.
Then, an object defined only on the $n$th overlaps ($n$-cochain)
will be identified as an $n$-form defined on the total space.
Note that this identification is consistent with the expected statistics.
For example, the fermionic ghost $b$ is identified as a $1$-form which is anticommuting.

The way we relate the (classical) BRST and Dolbeault/\v{C}ech cohomologies is as follows.
First, we embed both the BRST and Dolbeault cohomologies
to that of the combined operator $D+\deldol$.
Then, BRST and Dolbeault cohomologies are nothing but the special gauge choices
in the $D+\deldol$ cohomology, where non-minimal variables are absent (BRST),
and BRST ghosts are absent (Dolbeault).
Going back and forth between \v{C}ech and Dolbeault languages can be achieved
by imitating the standard argument in complex analysis,
i.e. by using a partition of unity to patch together \v{C}ech cochains
to obtain Dolbeault forms.
Although we will not explore in the main text,
it should be possible to directly relate the (minimal) BRST and \v{C}ech languages
by considering the cohomology of $D+\delcech$,
where $\delcech$ is the difference operator of \v{C}ech cohomology.

\bigskip
One of the virtue of studying these simpler models is that the BRST
description is very effective, allowing a close study of its
cohomology. In particular, the full partition function of the BRST
cohomology can be easily computed and it manifestly possesses two
important symmetries that we shall call ``field-antifield'' and
``$\ast$-conjugation'' symmetries. The former implies that, after
coupling to ``matter'' variables $(p_{i},\theta^{i})$, the
cohomology of the ``physical'' BRST operator $Q=\int
\lambda^{i}p_{i}$ comes in field-antifield pairs\footnote{This fact
and some topics related to our paper have been recently reported in
\cite{Chesterman:2008sq} for the simple model $N=2$.}. As such, the
symmetry is indispensable when one tries to define a sensible
``spacetime'' amplitudes.

The second symmetry, the $\ast$-conjugation symmetry, turns out to
be more powerful for analyzing the structure of the BRST
cohomology. It implies the existence of a non-degenerate inner
product that couples the cohomologies at ghost numbers $k$ and
$1-k$. In particular, there is a one-to-one mapping between
$H^{0}(D)$ and $H^{1}(D)$, and since $H^{k}(D)$ is empty for $k$
negative, all the higher cohomologies $H^{k}(D)$ with $k>1$ are
also empty. This ``vanishing theorem'' is rather important for the
pure spinor case ($H^{k}$ with $k>3$)~\cite{PS}.

\bigskip
The plan of this paper is as follows.
In section~\ref{sec:themodels}, after briefly reviewing
the general theory of the curved $\beta\gamma$ formalism,
we introduce the models to be considered in this paper,
both in curved $\beta\gamma$ and BRST descriptions.
As mentioned above, they are modeled after the ghost sector of the pure spinor superstring,
and the target spaces are simple cones defined by a single quadratic constraint.

In section~\ref{sec:partfn} we compare the partition functions of naive gauge invariant
polynomials and that of the BRST cohomology, and find that the latter includes
some extra states.
In fact, those ``extra'' states are essential for having field-antifield symmetry,
so perhaps it is more appropriate to refer to them as the states ``missing''
from the space of naive gauge invariant polynomials.

In section~\ref{sec:structBRST}, we study in detail the structure of
the quantum BRST cohomology.
It will be found that there is a one-to-one mapping between
the gauge invariant polynomials (elements of $H^{0}(D)$)
and the ``extra'' states (elements of $H^{1}(D)$).
Also, it will be shown that the quantum BRST cohomology
is empty outside those degrees.

Finally, in section~\ref{sec:equivcohoms},
the mapping between BRST and
\v{C}ech/Dolbeault curved $\beta\gamma$ descriptions is explained.
We shall show explicitly how the classical pieces of
the cohomology representatives are related
and point out how this correspondence can be broken
quantum mechanically.

An appendix is provided for explaining
some details of the curved $\beta\gamma$ description
of the models considered in this paper.

\section{The models}
\label{sec:themodels}

We begin with a brief review of the
basics of the theory of curved $\beta\gamma$ systems,
following~\cite{Malikov:1998dw}\cite{Kapustin:2005pt,Witten:2005px,Nekrasov:2005wg}.
The formalism is then used to introduce the models
by specializing the target space to be a simple quadric cone $\lambda^{i}\lambda^{i}=0$ ($i=1\sim N$).
The BRST descriptions of the same models are introduced in section~\ref{subsec:modelBRST},
and the geometries of the target space for some specific values of $N$
are explained in section~\ref{subsec:geomX}.

\subsection{Quick review of the curved $\beta\gamma$ formalism}

To construct a general curved $\beta\gamma$ system on a complex manifold $X$,
one usually starts with a set of free conformal field theories
taking values in the coordinate patches $\{U_A\}$ of $X$,
and tries to glue them together.
The field contents of each conformal field theory are
the (holomorphic) coordinate of a patch $u^{a}$
and its conjugate $v_{a}$ satisfying the
free field operator product expansion
\begin{align}
  \label{eq:localfreeOPE}
  u^{a}(z)v_{b}(w) &= {{\delta^{a}}_b \over z-w} \,.
\end{align}
Unlike the conventional sigma models on complex manifolds,
antiholomorphic coordinates need not be introduced.
On an overlap $U_{A}\cap U_{B}$,
two coordinates $u^{a}$ and $\Tilde{u}^{\Tilde{a}}$ are related
in the usual geometric manner,
\begin{align}
  \Tilde{u}^{\Tilde{a}} &= \Tilde{u}^{\Tilde{a}}(u)\,,
\end{align}
but it requires some thought to find
the gluing condition for the conjugates $v_a$ and $\Tilde{v}_{\Tilde{a}}$
because the classical relation,
\begin{align}
  \Tilde{v}_{\Tilde{a}} &\overset{?}{=}
   \tau_{\Tilde{a}}{}^{b}v_{b}\,,\quad
  \Bigl( \tau_{\Tilde{a}}{}^{b} = (\tau_{AB})_{\Tilde{a}}{}^{b} = {\del u^{b} \over \del \Tilde{u}^{\Tilde{a}}} \Bigr)\,,
\end{align}
suffers from normal ordering ambiguities.
In order to glue the free field operator products~(\ref{eq:localfreeOPE}) on an overlap,
the conjugates in two patches must be related as~\cite{Malikov:1998dw}\cite{Kapustin:2005pt,Witten:2005px,Nekrasov:2005wg}
\begin{align}
 \label{eq:vglue}
  \Tilde{v}_{a}
  &= \NO{ \tau_{a}{}^{b}v_{b}  }
    + \Tilde{\phi}_{\Tilde{a}\Tilde{b}}\del_z\Tilde{u}^{\Tilde{b}}\,,
\end{align}
where the correction $\Tilde{\phi}$ is a matrix defined on the overlap
and $\NO{\tau_{\Tilde{a}}{}^{b}v_{b}} =\NO{ ({ \del u^{b} / \del \Tilde{u}^{{\Tilde{a}}} })v_{b}  }$
denotes the usual free field normal ordering with respect to $u$ and $v$.
(There are no ordering ambiguities for $\Tilde{\phi}_{ab}\del_z\Tilde{u}^{b}$.)
It is convenient to decompose $\Tilde{\phi}$ into symmetric and antisymmetric pieces,
\begin{align}
 \Tilde{\phi}_{\Tilde{a}\Tilde{b}} = \Tilde{\sigma}_{\Tilde{a}\Tilde{b}} + \Tilde{\mu}_{\Tilde{a}\Tilde{b}} \,,
\end{align}
and regard the antisymmetric piece $\Tilde{\mu}_{\Tilde{a}\Tilde{b}}$ as the component of a two form
\begin{align}
\mu &= {1\over2}  \Tilde{\mu}_{ab}\targetd\Tilde{u}^{a}\wedge\targetd\Tilde{u}^{b} \,.
\end{align}
Solving $\Tilde{v}_{\Tilde{a}}(z)\Tilde{v}_{\Tilde{b}}(w)=0$,
one finds the conditions on $\Tilde{\sigma}$ and $\mu$ to be
\begin{align}
 \label{eq:vgluesol}
\begin{split}
\Tilde{\sigma}_{\Tilde{a}\Tilde{b}} &= -(\del_{c}\tau_{\Tilde{a}}{}^{d}\del_{d}\tau_{\Tilde{b}}{}^{c})
 =  - \Bigl( {\del^{2}u^{d} \over \del u^{c}\del\Tilde{u}^{a} }
   { \del^{2}u^{c} \over \del u^{d}\del\Tilde{u}^{b} } \Bigr)\,,
\\
 \targetd \mu
  &= -\tr( \tau^{-1}\targetd \tau)^3
  = -{\del\Tilde{u}^{\Tilde{a}} \over \del u^{b}}\targetd\Bigl( {\del u^{b} \over \del \Tilde{u}^{\Tilde{c}}}\Bigr)
   \wedge{\del\Tilde{u}^{\Tilde{c}} \over \del u^{d}}\targetd\Bigl({\del u^{d} \over \del \Tilde{u}^{\Tilde{e}}}\Bigr)
  \wedge{\del\Tilde{u}^{\Tilde{e}} \over \del u^{f} }\targetd\Bigl( {\del u^{f} \over \del \Tilde{u}^{\Tilde{g}}}\Bigr) \,.
\end{split}
\end{align}

The argument up to this point was local and the quantum correction $\phi\del u$
can always be found.
The $2$-form $\mu$ is the data assigned to every
double overlaps $U_{AB}=U_A\cap U_B$ so
it constitutes a \v{C}ech $1$-cochain; when we wish to emphasize this fact,
we denote $\mu=(\mu_{AB})$ etc.
Now, the solution to the gluing condition~(\ref{eq:vgluesol}) is not quite unique
and, at the same time, might not be compatible on the triple overlaps $U_{ABC}=U_A\cap U_B\cap U_C$.
The ambiguity comes from the freedom to add closed $2$-form valued \v{C}ech $1$-coboundaries to $\mu$
\begin{align}
  \mu=(\mu_{AB})\to \mu + \delcech \alpha = (\mu_{AB}+\alpha_A-\alpha_B)\,,\quad \alpha=(\alpha_A)\colon\text{closed $2$-form}\,,
\end{align}
which can be absorbed in the redefinitions of the local coordinates (and their conjugates)
in $U_A$ and $U_B$.
On the other hand,
the consistent gluing requires the following $2$-cocycle be a coboundary:
\begin{align}
\label{eq:anomaly2cocycle}
\begin{split}
  \psi &= (\psi_{ABC}) = \bigl(\mu_{AB}+\mu_{BC}+\mu_{CA}-\tr(\tau_{AB}\,\targetd\tau_{BC}\wedge\targetd\tau_{CA}) \bigr)\,.
\end{split}
\end{align}
In short, the moduli of the gluing is parameterized by
the first \v{C}ech cohomology $H^{1}(\calZ^2)$ of closed $2$-forms on $X$,
but it can be obstructed by the second cohomology $H^{2}(\calZ^2)$
(or the first Pontryagin class $p_1(X)$).
Also, the gluing of the global symmetry currents of $X$
are parameterized and possibly obstructed by similar (``equivariant version'' of)
cohomologies.

Finally, let us recall that even if the operator products~(\ref{eq:localfreeOPE})
and the symmetry currents could be consistently glued,
one may not be able to define the energy-momentum tensor $T$ globally.
This implies the violation of the conformal symmetry due to an anomaly.
For $T$ to be well-defined, one has to improve it using a nowhere vanishing holomorphic top
form $\Omega$ of $X$.
The obstruction to having $\Omega$ is the first Chern class $c_1(X)$.
Hence to have a globally defined conformal field theory as a curved $\beta\gamma$ system,
the target $X$ must be a Calabi-Yau space (though strictly speaking $X$ need not be K\"{a}hler).

This concludes our brief review of the basic notions of
the theory of curved $\beta\gamma$ systems.

\subsection{Curved $\beta\gamma$ description}
\label{subsec:modelsBC}

From now on, we specialize to a subset of curved $\beta\gamma$ systems where
the target space $X$ is a cone in $\mathbb{C}^{N}$
defined by a quadratic constraint~\cite{Grassi:2006wh}
\begin{align}
 \label{eq:GeneralConstraint}
X &= \{ \lambda^{i}\in \mathbb{C}^{N}\;|\;
  G\equiv \lambda^{i}\gamma_{ij}\lambda^{j} = 0\,,\quad \lambda\ne0 \}\,,\quad(i\,,j = 1\sim N)\,.
\end{align}
Here, $\gamma_{ij}$ is some non-degenerate symmetric constant ``metric''.
Of course, one can always diagonalize as $\gamma_{ij}=\delta_{ij}$
so we drop the factor of $\gamma$ and its inverse,
and do not distinguish upper and lower indices.

Since we remove the origin $\lambda=0$ as indicated above,
$X$ is a $\mathbb{C}^{\ast}$-bundle over the base $B=X/\mathbb{C}^{\ast}$
where the quotient acts by the global rescaling of $\lambda$.
The target space reparameterization (Pontryagin) anomaly is absent
just as in the pure spinor case.
That is, although the base $B$ has
a non-trivial anomaly $2$-cocycle $\psi$~(\ref{eq:anomaly2cocycle}),
its extension to the total space $X$
represents a trivial \v{C}ech class
by virtue of the fiber direction
(see appendix~\ref{app:geomcone})~\cite{Nekrasov:2005wg}.
Therefore, the conjugate $\omega_{i}$, or more precisely its independent components, can be glued consistently.
For the case at hand, the symmetry currents for the $SO(N)$ rotation $N_{ij}$
and rescaling of the cone $J$ can also be defined consistently\footnote{%
$J$ is often called as ``ghost number'' current in the literature.
But we shall call it ``$t$-charge current'' instead to avoid the confusion with
the BRST ghost number introduced later.}.
Furthermore, $X$ is a (non-compact) Calabi-Yau space admitting a nowhere vanishing
holomorphic top form.
Thus, the energy-momentum tensor can be globally defined
and the curved $\beta\gamma$ theory on $X$ is conformally invariant.

For completeness, we give in appendix~\ref{app:geomcone}
some more details of the curved $\beta\gamma$ description
such as the choice of local coordinates
and the expressions of the currents $(J,N,T)$ etc.

\paragraph{Non-minimal or Dolbeault description}

When dealing with operators that are not globally defined,
it is notationally
more convenient to introduce the non-minimal variables defined as~\cite{Berkovits:2005bt}
\begin{align}
\begin{split}
  \overline{\lambda}_{i}\,,\; \overline{\omega}^{i}\,,
  &\quad\bigl(\overline{\lambda}_{i}\overline{\lambda}_{i}=0\,,\quad
   \delta_{\Lambda,\Psi}\overline{\omega}^{i} = \Lambda\overline{\lambda}^{i} + \Psi r^{i} \bigr)\,, \\
  r_{i}=\targetd\overline{\lambda}_{i}\,,\; s^{i}\,,
  &\quad\bigl(r_{i}\overline{\lambda}_{i} = 0\,,\quad \delta_{\Lambda}s^{i} = \Lambda\overline{\lambda}^{i}\bigr)\,.
\end{split}
\end{align}
Observe that $\overline{\lambda}$ satisfies the same constraint as $\lambda$.
In the language of complex geometry, $\overline{\lambda}$ corresponds to the antiholomorphic
coordinate of the target space $X$.
The virtue of introducing those extra variables is that one can deal with
globally defined operators, often hiding the explicit dependence on the local coordinates.
The mapping between \v{C}ech and Dolbeault descriptions can be explicitly done
using the partition of unity given in appendix~\ref{app:geomcone}.

Physical states in non-minimal formalism are defined as the cohomology of the Dolbeault operator
\begin{align}
  \deldol = -r_{i}\overline{\omega}^{i} \sim {\targetd\overline{\lambda}_{i}}{\del\over \del\overline{\lambda}_{i}} \,.
\end{align}
If one wishes to be rigorous, this gauge invariant expression
should be understood in terms of the local coordinates.
Despite its simple form, the cohomology of $\deldol$ is not quite trivial,
because the minimal variables are constrained,
and because one allows poles in $(\lambda\overline{\lambda})$.

However, non-zero modes of the non-minimal variables do not affect the cohomology
due to the relation
\begin{align}
  \deldol(s\del\overline{\lambda}) = \overline{\omega}\del\overline{\lambda} + s\del r = -T_{\text{non-min}} \,.
\end{align}
Whenever there is a $\deldol$-closed operator $F$
with positive weight $h$ carried by the non-minimal sector,
it is also a $\deldol$ of itself
multiplied by the zero-mode of $s\del\overline{\lambda}$:
\begin{align}
  -{1\over h}\deldol\bigl( (s\del\overline{\lambda})_{0}F\bigr) = F \,.
\end{align}
Similarly, due to the relation
\begin{align}
  \deldol(s\overline{\lambda})
 = \overline{\omega}\overline{\lambda} + s r = -J_{\text{non-min}} \,,
\end{align}
the zero-modes of $\overline{\lambda}$ and $r$ can only appear in
the non-minimal charge $0$ combinations
\begin{align}
 (\lambda\overline{\lambda})^{-1}{\overline{\lambda}_{i}}
\quad\text{and}\quad
 ({\lambda\overline{\lambda}})^{-1}{r_{i} }\,.
\end{align}
Given those restrictions on the appearance of non-minimal variables,
it follows that whether they are constrained or not
is irrelevant for the cohomology of $\deldol$.
That is, even if one regards the non-minimal variables
as {\em unconstrained},
the cohomology of $\deldol$ remains unchanged.

\paragraph{``Gauge invariance'' in curved $\beta\gamma$ framework}

When discussing the constrained curved $\beta\gamma$ systems above,
we used the notion of ``gauge invariance'' to define the space
on which the \v{C}ech or Dolbeault operators act.
In the curved $\beta\gamma$ framework, however, one usually deals with the operators
defined intrinsically on the target space $X$ (even in the non-minimal language),
and does not worry about the ``gauge invariance''.
Let us explain the relation between the two descriptions.

For simplicity, consider the particle moving on the cone $X$.
When one speaks of the gauge transformation $\delta_{\Lambda}\omega_{i}=\Lambda\lambda_{i}$,
it is implicitly assumed that the phase space $T^{\ast}X$
is embedded in a Euclidean space $(\omega,\lambda)\in T^{\ast}\mathbb{C}^{N}=\mathbb{C}^{2N}$.
Then, a gauge transformation generates a motion vertical to $T^{\ast}X$,
and the gauge invariance of an object simply means that it is living inside $T^{\ast}X$.
In the curved $\beta\gamma$ language, $T^{\ast}X$ is treated intrinsically
and everything is manifestly gauge invariant;
there is really no way to construct ``gauge {\em non}-invariant object''
just by using the local coordinates on $T^{\ast}X$.
Therefore, ``gauge invariance'' is a convenient way to refer to the operators
defined intrinsically on $X$, but by using the ``extrinsic'' coordinates $(\omega,\lambda)$.

Note, however, that the converse is not necessarily true.
For example, there can be operators that are globally defined on $X$,
but nevertheless cannot be described as a gauge invariant polynomial
in $(\omega,\lambda;\del)$.
For the class of models considered in the present paper,
this will happen when the dimension
of $X$ is smaller than $3$ ($N<4$),
i.e. when the base $B$ of $X$ has one or ``zero'' dimensions.

The reason why we find it useful to work in the space of
``gauge invariant'' functions of $(\omega,\lambda;\del)$ is the following.
Later in section~\ref{sec:equivcohoms} we explore the relation
between the BRST and the intrinsic curved $\beta\gamma$ descriptions of the constraints.
Since $(\omega,\lambda)$ are promoted to genuine free fields in the BRST framework,
what naturally appears there is $X$ embedded in a flat space,
rather than its intrinsic description.

\subsection{BRST description}
\label{subsec:modelBRST}

For the model with the irreducible constraint~(\ref{eq:GeneralConstraint})
the conventional BRST formalism provides a very simple way of describing it,
compared to the elaborate language of the curved $\beta\gamma$ formulation.
(This is not necessarily the case for
infinitely reducible constraints such as the ones for the pure spinors.)
Here, a fermionic $(b,c)$ ghost pair is introduced to impose the constraint effectively
and the physical states are described as the cohomology of the BRST operator
\begin{align}
  D &= \int b(\lambda\lambda)\,.
\end{align}
The ghost number $0$ cohomology $H^{0}(D)$
reproduces the space of globally defined gauge invariant polynomials.
However, there are also non-trivial cohomologies at non-zero ghost numbers.
Typical example is the ghost $b$ itself with ghost number $+1$.
Describing explicitly the operator corresponding to $b$ in the curved $\beta\gamma$
language is one of the goals of the present paper.

As will be shown in section~\ref{sec:structBRST},
the cohomology turns out to be non-vanishing
only at ghost numbers $0$ and $1$.
Moreover, we find that the elements of $H^{0}(D)$ and $H^{1}(D)$
are paired under a certain inner product.

We expect that this property of the BRST cohomology is
a general property of the theory defined by a system of homogeneous constraints.
In particular, for the important case of the pure spinor model
the non-vanishing cohomologies are $H^{0}(D)$ and $H^{3}(D)$,
which again come in pairs~\cite{PS}.

\subsection{Geometries of $X$ and models with lower values of $N$}
\label{subsec:geomX}

In the forthcoming sections we will assume $N\ge4$
and our discussions will not depend on the specific value of $N$.
One can define consistent models for $2\le N\le3$
both in the BRST and the curved $\beta\gamma$ frameworks,
but the two descriptions will not be equivalent even classically
(at least when they are defined analogously to the $N\ge4$ case).
Here, we explain the geometry of $X$ for some values of $N$,
and give a rationale behind the restriction on $N$.
Appendix~\ref{app:geomcone} provides some additional properties of $X$
such as coordinate systems and the associated partition of unity etc.

\paragraph[$\lambda^2=0$ ($N=1$)]{\boldmath $\lambda^2=0$ ($N=1$)}

The model with $N=1$ has a single coordinate variable $\lambda$
which is constrained as $\lambda^{2} = 0$.
As such, the ``target space'' is not geometrical in the usual sense,
and it is not clear what local coordinates one should take
to define the curved $\beta\gamma$ model intrinsically.

Also on the BRST side, this model is qualitatively different
from $N\ge2$ models because
$\lambda^{2}$ and its derivatives $\del^{n}\lambda^{2}$ are not independent.
Therefore, the naive BRST charge $D =\int b\lambda^2$ has
extra cohomologies outside $H^{0}(D)$ and $H^{1}(D)$.
By appropriately introducing a chain of ghosts-for-ghosts,
one should be able to describe the
gauge invariant polynomials in $(\omega,\lambda)$ as the zeroth cohomology.
But let us avoid this effort in the present paper,
since we explain the BRST construction for
the reducible pure spinor constraints in detail in~\cite{PS}.
Instead, we explicitly identify some unwanted cohomology elements
in section~\ref{subsec:remlam2}.

\paragraph[$\lambda\Tilde{\lambda}=0$ ($N=2$)]{\boldmath $\lambda\Tilde{\lambda}=0$ ($N=2$)}

For $N\ge2$, the constraint $\lambda^{i}\lambda^{i}=0$ is irreducible
so the BRST operator should be given by $D=\int b(\lambda\lambda)$
and the structure of its cohomology does not depend on $N$.
Also, the space $X$ defined by the constraint
is non-degenerate and the curved $\beta\gamma$ system on $X$ is consistent.
However, for $N=2,3$ models
the two descriptions do not agree because
the intrinsic curved $\beta\gamma$ description allows
some (globally defined) operators which cannot be
described as the polynomials of the
extrinsic coordinates $(\omega,\lambda)$.

Defining
\begin{align}
  (\lambda,\Tilde{\lambda})= (\lambda^{1}+\mathi \lambda^{2},\lambda^{1}-\mathi \lambda^{2})\,,
\end{align}
the constraint for the $N=2$ model can be expressed as
$\lambda\Tilde{\lambda}=0$. So the geometry of $X$ is a simple
cone, but it becomes a union of two disjoint components when the
origin is removed. As such, the intrinsic description of the
system on $X$ is quite different from its embedding in the flat
space $(\omega,\lambda)$, and hence from the BRST description.
(The BRST treatment of this model and the enumeration of gauge
invariant polynomials up to level $2$ was studied
in~\cite{Grassi:2006wh}.)

\paragraph[$xy-z^2=0$ ($N=3$)]{\boldmath $xy-z^2=0$ ($N=3$)}

Similarly, the constraint for the $N=3$ model can be rephrased
as $xy-z^{2}=0$ where the new variables are defined as
\begin{align}
  (x,y,z) &= (\lambda^{1}+\mathi \lambda^{2},\lambda^{1}-\mathi \lambda^{2},\mathi \lambda^{3})\,.
\end{align}
The space $X$ is in fact a simple singular Calabi-Yau space
\begin{align}
  \mathbb{C}^{2}/\mathbb{Z}^{2}\,,
\end{align}
which has a so-called $A_1$ singularity at the origin $(x,y,z)=(0,0,0)$.
This can be seen by using the coordinate $(a,b)\in\mathbb{C}^{2}$
and defining
\begin{align}
  (x,y,z) = (a^2,b^2,ab)\,.
\end{align}
The division by $\mathbb{Z}^{2}$ identifies
a point $(a,b)$ with the antipodal point $(-a,-b)$.

Although the curved $\beta\gamma$ system on $X$ by itself is perfectly sensible,
it is not equivalent even classically to the BRST system with $D=\int b(xy-z^2)$.
The reason is because, at a given mass level,
there is a finite number of globally defined operators on $X$
that cannot be written as gauge invariant polynomials in $(x,y,z)$.
For example, as noted in~\cite{Grassi:2007va},
there is one such operator at the first mass level.
In the coordinate system $(g,u)\in U_1$ and $(\Tilde{g},\Tilde{u})\in U_{\Tilde{1}}$
($X=U_1\cup U_{\Tilde{1}}$)
given in appendix~\ref{app:geomcone},
the extra state is given by the \v{C}ech $0$-cocycle
\begin{align}
 F = (F_{1}, F_{\Tilde{1}})
  =(g\del u, -\Tilde{g}\del{\Tilde{u}})\,,\qquad\bigl(g=\Tilde{g}\Tilde{u}^2\,,\; u=1/\Tilde{u} \bigr) \,.
\end{align}
The coordinate patches $U_1$ and $U_{\Tilde{1}}$ correspond
to the region $x\ne0$ and $y\ne0$ respectively,
and $F$ can also be expressed as~\cite{Grassi:2007va}
\begin{align}
 \label{eq:nonpolyF}
 F = (z x^{-1}{\del x} - \del z , -z y^{-1}{\del y} + \del z) \,.
\end{align}
Clearly, there is no corresponding operator
in the BRST cohomology computed in the polynomial regime,
so the $N=3$ curved $\beta\gamma$ model is different from
the BRST model.

One might worry if there exist non-trivial elements of the \v{C}ech cohomology
for $N\ge4$ models $xy-z_a z_a=0$ ($a=3\sim N$) as well,
but it can be argued that there are none.
Note that the existence of $F$ crucially depends on the fact that the
base $B$ of $X$ is one dimensional.
In higher dimensions ($N\ge4$),
the angular coordinate $u_{a}\in B$ carries an index
and transforms like $u_{a} = \Tilde{u}_{a}(\Tilde{u}\cdot\Tilde{u})^{-1}$.
So $\del u_{a} \in U_1$ have
a pole $(\Tilde{u}\cdot\Tilde{u})^{-2}$ in another patch $U_{\Tilde{1}}$,
and the only way to cancel the pole is to multiply it by
$g^2=\Tilde{g}^{2}(\Tilde{u}\cdot\Tilde{u})^{2}$.
But $g^{2}\del u_{a}$ (unlike $g\del u_{a}$)
is in fact a polynomial $z_{a}\del x - x\del z_{a}$.
Similarly, there should be no non-polynomial operators at higher mass levels.

Another way to understand this is to note that $F$ in~(\ref{eq:nonpolyF}) does not have
a corresponding operator on a slightly deformed space $X_{\epsilon}\colon xy-z^{2}=\epsilon$.
That is, the order $\epsilon$ term of the deformed operator $F_{\epsilon}=F+\epsilon F'+\cdots$ has a pole in $z$
and hence is not globally defined on $X_{\epsilon}$.
Therefore, for $N\ge4$ where the additional coordinates $\lambda^{i}$ ($i=4\sim N$)
can play the role of $\epsilon$, there will not be the extra operators
analogous to $F$.

For those reasons, we assert for $N\ge4$ models
that all the elements of the \v{C}ech cohomology
can be represented using the extrinsic coordinates $(\omega,\lambda)$,
though we do not have a rigorous proof.

\paragraph[$xy-zw=0$ ($N=4$)]{\boldmath $xy-zw=0$ ($N=4$)}

The target space of the $N=4$ model is the famous conifold
as can be seen from defining
\begin{align}
  (x,y,z,w) &= (\lambda^{1}+\mathi \lambda^{2},\lambda^{1}-\mathi \lambda^{2},\mathi\lambda^{3}-\lambda^{4},\mathi \lambda^{3}+\lambda^{4})\,.
\end{align}
(A partial enumeration of gauge invariant polynomials
up to level $2$ for this model
was studied in~\cite{Grassi:2007va}.)

As explained above, all
the models with $N\ge4$ should behave qualitatively the same.
In particular, we shall argue that its curved $\beta\gamma$ description is
equivalent to the BRST description.

\paragraph[$D=8$ pure spinor ($N=8$)]{\boldmath $D=8$ pure spinor ($N=8$)}

We have been implicitly assuming that $\lambda^{i}$ transforms as a vector of $SO(N)$.
For the special value of $N=8$, however, $\lambda^{i}$ is not significantly different from
the $SO(8)$ (chiral) spinor $\lambda^{a}$ due to the triality.
$\lambda^{a}$ satisfying $\lambda^{a}\lambda^{a}=0$
is in fact nothing but the Cartan pure spinor in eight dimensions.

\section{Partition function, its symmetries and the extra states}
\label{sec:partfn}

As mentioned in the introduction, the main motivation for the present investigation is
to understand the proper Hilbert space for the pure spinor superstring
in a simplified setup.
We begin the study by computing the partition function of
the gauge invariant polynomials, by explicitly counting
them at several lower mass levels.
Our main finding will be that, starting from the first mass level,
the space of naive gauge invariants lacks the field-antifield symmetry
because of some finite number {\em fermionic} operators that are missing.

On the contrary, the partition function of the BRST cohomology
is found to enjoy the field-antifield symmetry.
Since the ghost number $0$ sector of the BRST cohomology is
(classically) equivalent to
the space of gauge invariant polynomials, this means that the states
depending essentially on the ghosts are very important.
Those extra states are explicitly identified in section~\ref{subsec:extraBRST}.

Also, the BRST partition function is found to
possess another discrete symmetry
which we call ``$\ast$-conjugation symmetry''.
Both field-antifield and $\ast$-conjugation symmetries
reflect certain dualities of the cohomology,
and their existence plays an important role
for the consistency of the pure spinor formalism.

\subsection{Definition of the partition function}

We begin by describing the definition of our partition function.
The characters of the states we are interested in are
\begin{itemize}
\item statistics (Grassmanity) measured by $(-1)^F$ ($F$: fermion number operator),
\item weight (Virasoro level) measured by $L_{0}$, and
\item $t$-charge measured by a $U(1)$ charge $J_{0}$.
\end{itemize}
By introducing formal variables $(q,t)$ to keep track of the charges,
the partition function is defined as
\begin{align}
 Z(q,t) &= \Tr_{\calH}(-1)^{F}q^{L_{0}}t^{J_{0}} \,.
\end{align}
What we are really interested in is the Hilbert space $\calH$
in which the trace is taken over,
and we shall define the currents for $L_0$ and $J_0$ in the next paragraph.

In the BRST framework, basic fields obey free field operator products,
and the ghost extended
energy-momentum tensor and the $t$-charge current are defined as
\begin{align}
  T &= -\omega_{i}\del\lambda^{i} - b\del c \,,\quad
  J = -\omega_{i}\lambda^{i} - 2 bc \,.
\end{align}
The charges of the basic operators are
\begin{align}
\begin{split}
  F(\omega,\lambda)&=(0,0)\,,\quad h(\omega,\lambda) = (1,0)\,,\quad t(\omega,\lambda) = (-1,1)\,, \\
  F(b,c)&=(1,1)\,,\quad h(b,c)=(1,0)\,,\quad t(b,c)=(-2,2)\,.
\end{split}
\end{align}
In particular, the BRST operator $D=\int b(\lambda\lambda)$
is neutral both under $L_{0}$ and $J_{0}$, so the
partition function of $D$-cohomology is insensitive to
quantum corrections.
(Similar remark applies for the \v{C}ech/Dolbeault cohomologies
for the intrinsic description.)

Let us remark in passing that we define the ghost number current as
\begin{align}
  J_g &= +bc\,,
\end{align}
so that the ghost numbers are
\begin{align}
  g(b,c;D) &= (1,-1;1)\,.
\end{align}

In the curved $\beta\gamma$ framework, construction of $T$ and $J$ are more complicated
but their existence is assured by the general theory
as we briefly recalled above~\cite{Malikov:1998dw}\cite{Kapustin:2005pt,Witten:2005px,Nekrasov:2005wg}.
Their explicit expressions are given in appendix~\ref{app:geomcone} for completeness.
Here, let us only mention that they can be constructed and that
the $t$-charges of operators can be correctly inferred by expressing them
in terms of the ``extrinsic coordinates'' $(\omega,\lambda)$ carrying $t$-charges $(-1,1)$.
For example, the $t$-charge of $J=-\omega\lambda+(\text{quantum corrections})$ itself is $0$.

\subsection{Gauge invariant polynomials}
\label{subsec:GIpoly}

We now count the number of gauge invariant polynomials
constructed out of $\lambda$, $\omega$ and their derivatives,
and compute the partition function $Z(q,h) = \Tr (-1)^{F}q^{L_{0}}t^{J_{0}}$.
(Similar counting of gauge invariant polynomials
for the present and related models
is given in~\cite{Grassi:2007va}.)

\paragraph{Weight $0$}

At the lowest level, the states are exhausted by
\begin{align}
  \lambda^{\lpar i_{1}} \cdots \lambda^{i_n \rpar}\,.
\end{align}
Here, the notation $\lpar i_1\cdots i_n \rpar$
signifies the symmetric traceless tensor product.
The states can be conveniently described
using the Dynkin labels for $SO(N)\times U(1)_t$ as
\begin{align}
  \dynkin(n00\cdots0)t^{n}\,.
\end{align}
Using the well-known dimension formulas for the symmetric tensors\footnote{%
Strictly speaking, those formulas are correct only for $k\ge2$.
Dimensions of symmetric traceless tensors for $N=2,3$ are
$SO(2)=2-\delta_{n,0}$ and $SO(3)=2n+1$.}
\begin{align}
\dim\dynkin(n00\cdots0) &= \begin{cases}
 \prod_{i=2}^{k}{(n + 2k - i - 1)(n + i - 1) \over (2k - i - 1)(i - 1)} & SO(2k)\,,\\
 \prod_{i=2}^{k}{(n-2k+1)(n+i-1) \over (2k-1)(i-1)} & SO(2k+1)\,,
\end{cases}
\end{align}
one gets
\begin{align}
 \label{eq:Z0gi}
  Z_{0}(t) &= \sum_{n=0}^{\infty}\dim\dynkin(n00\cdots0)t^{n}
  = {1-t^{2} \over (1-t)^{N}} \,.
\end{align}

Note that the level $0$ partition function is invariant under
\begin{align}
\text{``field-antifield symmetry''}\colon\quad Z_{0}(t) &= -(-t)^{2-N}Z_{0}(1/t)\,.
\end{align}
As explained in~\cite{Berkovits:2005hy},
the number $2-N$ on the exponent is the ghost number anomaly of the system.
Since this symmetry plays an important role in our forthcoming discussions
(as well as in the pure spinor superstring),
let us explain the implication of its existence
before going on to the weight $1$ partition function.

\paragraph{Field-antifield symmetry}

Suppose one couples the system to free fermionic $bc$ systems
$(p_{i},\theta^{i})_{i=1\sim N}$ of weight $(1,0)$,
and extends the definition of the $t$-charge
to the new sector as $t(p,\theta)=(-1,1)$.
By an analogy with the pure spinor superstring,
one also defines
the ``physical'' BRST operator as
\begin{align}
Q=\int \lambda^{i}p_{i} \,.
\end{align}
Then the symmetry $Z_{0}(t) = -(-t)^{2-N}Z_{0}(1/t)$ implies
that all $Q$-cohomology elements appear
in ``spacetime'' field-antifield pairs
\begin{align}
  \text{$V$ \ at \ $\pm t^{n}$}\quad\leftrightarrow\quad
  \text{$V_A$ \ at \ $\mp t^{2-n}$}\,.
\end{align}
Indeed, the total zero-mode partition function reads
\begin{align}
  \mathbf{Z}_{0}(t) = Z_{\lambda,0}(t)Z_{\theta,0}(t) = 1- t^{2}\,,
\end{align}
which is accounted for by a pair of ``massless'' cohomologies
\begin{align}
  \text{$1$ \ at \ $t^{0}$}\quad \leftrightarrow \quad \text{$(\lambda\theta)=\lambda^{i}\theta^{i}$ \ at \ $-t^{2}$}\,.
\end{align}

The field-antifield symmetry
implies the existence of a non-degenerate inner product that pairs every operator $V$
to its antifield $V_A$
\begin{align}
  (V,V_A) &= 1\,.
\end{align}
For the case at hand, the inner product can be defined as the overlap
\begin{align}
 (V,W)&=
\lim_{z\to0}\bra<0|z^{2L_0}V(1/z)W(z)\ket|0>\,,
\end{align}
with the condition
\begin{align}
  \bra<0|(\lambda\theta)\ket|0> = 1\,.
\end{align}
It is easy to see that $Q$-exact states decouples from the inner product.
Of course, this construction of the inner product is reminiscent of that
of the pure spinor superstring~\cite{Berkovits:2000fe}
where one uses the rule
\begin{align}
  \bra<0| (\lambda\gamma^{\mu}\theta)(\lambda\gamma^{\nu}\theta)(\lambda\gamma^{\rho}\theta)(\theta\gamma_{\mu\nu\rho}\theta) \ket|0> = 1\,.
\end{align}

We will shortly observe that the space of gauge invariant polynomials at weight $1$
and higher lacks the field-antifield symmetry.
It might sound harmless but
we stress the importance of having the field-antifield symmetry at all
mass levels to define the ``spacetime amplitude'' appropriately.
Otherwise, some ``massive'' vertex operators in the cohomology of $Q=\int \lambda^{i}p_{i}$
would unfavorably decouple from the amplitude.
In fact, in the pure spinor formulation of superstring,
demonstrating the existence of field-antifield symmetry
for the full cohomology of $Q=\int \lambda^{\alpha}d_{\alpha}$ was an unresolved challenge.
This and related issues will be reported in a separate communication~\cite{PS}.

\paragraph{Weight $1$}

Having explained the notion of field-antifield symmetry,
let us go back to the construction of gauge invariant polynomials at weight $1$.
Here, one of $\del \lambda$ or $\omega$ can be used to saturate the weight.
$\del \lambda$ must satisfy the constraint at level $1$,
$\del(\lambda\lambda) = 2\lambda\del \lambda=0$,
while the conjugate $\omega$ must appear in the combination which is
invariant under the gauge transformation $\delta_{\Lambda}\omega = \Lambda \lambda$.
At level $1$, this condition implies that $\omega$ must appear in the form of
the gauge invariant currents $J$ and $N_{ij}$.
Hence, the gauge invariant polynomials are ($n\ge0$)
\begin{align}
\label{eq:h1gi}
\begin{split}
 \del \lambda^{\lpar j}\lambda^{i_1} \cdots \lambda^{i_n \rpar} &= \dynkin(n+1,00\cdots0)t^{n+1}\,, \\
 \del \lambda^{[j} \lambda^{\lpar k]}\lambda^{i_{1}}\cdots \lambda^{i_n \rpar} &= \dynkin(n10\cdots0)t^{n+2}\,, \\
 \omega_{j}\lambda^{\lpar j}\lambda^{i_{1}}\cdots \lambda^{i_{n} \rpar}&= \dynkin(n00\cdots0)t^{n}\,,\\
 \omega^{[j}\lambda^{\lpar k]}\lambda^{i_{1}}\cdots \lambda^{i_{n} \rpar}&= \dynkin(n10\cdots0)t^{n}\,.
\end{split}
\end{align}
Summing up the dimensions as before, one finds
\begin{align}
\label{eq:Z1poly}
\begin{split}
  Z_{1,\text{poly}}(t)
&= {Nt - t^2 - Nt^3 + t^4 \over (1-t)^{N}}+
{ \bigl(-1+(1-t)^{N} \bigr)t^{-2} + Nt^{-1} + 1 - Nt \over (1-t)^{N}} \,.
\end{split}
\end{align}
The first term represents the contribution from $\del\lambda$
and the second term represents that of $\omega$.

Note that $Z_{1,\text{poly}}(t)$ as defined in~(\ref{eq:Z1poly}) does not posses
the field-antifield symmetry.
However, it is easy to see from the way we wrote it that
\begin{align}
Z_1(t) &=   Z_{1,\text{poly}}(t) - t^{-2}
\end{align}
satisfies the symmetry.
This suggests that one needs an extra {\em fermionic} state with $t$-charge $-2$.
In the BRST cohomology, this extra state corresponds to the ghost $b$.
At first sight, there seems to be no room for fermionic states in the present setup,
but in fact they can be employed as the elements of
\v{C}ech-Dolbeault cohomologies at odd degrees.

\paragraph{Weight $2$}

Explicit constructions of the gauge invariant polynomials goes the same at
the level $2$.

First, there are polynomials with two $\omega$'s ($n\ge0$):
\begin{align}
\label{eq:h2w2}
\begin{split}
N_{\lbra i_1i_2}N_{i_3i_4\rbra}\lambda^{(n)}
 &= (\delta_{j_1\lbra i_1}\omega_{i_2})(\omega_{i_3}\delta_{i_4\rbra j_2})\lambda^{\lpar j_1}\lambda^{j_2}\lambda^{k_1} \cdots \lambda^{k_n \rpar}
  = \dynkin(n200\cdots0)t^{n} \,,\\
N_{i_0i_1}N^{i_0i_2}\lambda^{(n)}
  &=(\delta_{i_0j_1}\omega_{\lbra i_1})(\delta^{i_0j_2}\omega_{i_2\rbra})\lambda^{\lpar j_1}\lambda^{j_2}\lambda^{k_1} \cdots \lambda^{k_n \rpar}
  = \dynkin(n+2,00\cdots0)t^{n}\,,
 \\
N_{i_1i_2}J\lambda^{(n)}
 &= (\delta_{j_1[i_1}\omega_{i2]})\omega_{j_2}\lambda^{\lpar j_1}\lambda^{j_2}\lambda^{k_1} \cdots \lambda^{k_n \rpar}
  = \dynkin(n100\cdots0)t^{n}\,,\\
JJ\lambda^{(n)}
 &= \omega_{j_1}\omega_{j_2}\lambda^{\lpar j_1}\lambda^{j_2}\lambda^{k_1} \cdots \lambda^{k_n \rpar}
  = \dynkin(n00\cdots0)t^{n}\,.
\end{split}
\end{align}
Here, the symbol $\lbra i_1i_2\cdots i_n\rbra$ implies that the indices are traceless, block-symmetric,
and antisymmetric within each blocks; in particular $\lbra i_1,i_2 \rbra$
simply denotes the traceless symmetric tensor.

Also, there is a gauge invariant function with negative $t$-charge:
\begin{align}
\begin{split}
  f_{i} &= J\omega_{i} + N_{ij}\omega^{j} \\
  &= - 2(\lambda\omega)\omega_{i} + (\omega\omega)\lambda_{i}\,.
\end{split}
\end{align}
In a local coordinate patch $U_{1}=(g,u_{a})$,
components of $f_{i}$ are given by $(v_{a}v_{a}) / g$ and
its Lorentz transformations,
both classically and quantum mechanically.
Note, however, that polynomials of the form
$f_{i}\,\lambda^{(n+1)}$ ($n\ge0$) are not independent from
the ones listed in~(\ref{eq:h2w2}).

As for the polynomials with a single derivative and a single $\omega$,
one finds the following independent states ($n\ge0$):
\begin{align}
\label{eq:h2wdel}
\begin{split}
 N^{ij}\del\lambda \lambda^{(n)}
&= \bigl( \omega^{[i}\lambda^{\lpar j]}\del\lambda^{k}\lambda^{k_1}\cdots \lambda^{k_n\rpar}
  +\omega^{[i}\del\lambda^{k}\lambda^{\lpar j]}\lambda^{k_1}\cdots \lambda^{k_n\rpar} \\
&\qquad +\del\lambda_{i}\omega^{[i}\lambda^{\lpar j]}\lambda^{k_1}\cdots \lambda^{k_n\rpar}\bigr)
  + \del\lambda_{[k}\delta_{\ell]m}\omega^{[i}\lambda^{\lpar j]}\lambda^{m}\lambda^{k_1}\cdots \lambda^{k_n\rpar}
\\
&= \bigl( \dynkin(n+1,10\cdots)+\dynkin(n010\cdots)+\dynkin(n+1,0\cdots) \bigr)t^{n+1}
 + \dynkin(n20\cdots)t^{n+2}\,,
\\
J\del\lambda\lambda^{(n)}
 &= \omega_{j}\del\lambda^{\lpar i}\lambda^{j}\lambda^{k_1}\cdots \lambda^{k_n\rpar}
  + \omega_{k}\del\lambda^{[i}\lambda^{\lpar j]}\lambda^{k}\lambda^{k_1}\cdots \lambda^{k_n\rpar}\\
 &= \dynkin(n+1,0\cdots)t^{n+1}+\dynkin(n10\cdots)t^{n+2} \,, \\
T &= \omega_{i}\del\lambda^{i} = \dynkin(00\cdots)t^{0}\,.
\end{split}
\end{align}
Note that we could have included the energy momentum tensor $T$
as the ``$n=-1$ piece'' of the $J\del\lambda\lambda^{(n)}$ series;
in other words, $T\lambda^{(n+1)}$ and $J\del\lambda\lambda^{(n)}$ ($n\ge0$) are not
independent.

Finally, there are two types of polynomials with two derivatives,
$\del^{2}\lambda\lambda^{(n)}$ and $(\del\lambda)^{2}\lambda^{(n)}$,
but some of them are related by the level $2$ constraint
\begin{align}
  \lambda\del^{2}\lambda + \del\lambda\del\lambda = 0\,.
\end{align}
A choice of independent polynomials are ($n\ge0$):
\begin{align}
\label{eq:h2del2}
\begin{split}
\del^{2}\lambda^{i}\lambda^{\lpar j_1} \cdots \lambda^{j_n \rpar}
 &= \dynkin(10\cdots0)\otimes\dynkin(n0\cdots0)t^{n+1}\,, \\
\del\lambda^{\lpar i_1}\del\lambda^{i_2}\lambda^{j_1}\cdots \lambda^{j_n\rpar}
 &= \dynkin(n+2,0\cdots0)t^{n+2}\,, \\
\del\lambda^{[i_1}\lambda^{\lpar j_1]}\del\lambda^{j_2}\lambda^{j_2}\cdots \lambda^{j_n\rpar}
 &=\dynkin(n+1,10\cdots0)t^{n+3}\,, \\
 (\del\lambda^{\lbra i_1}\delta^{j_1}_{k_1}) (\del\lambda^{i_2}\delta^{j_2\rbra}_{k_2})\lambda^{\lpar k_1}\cdots \lambda^{k_n\rpar}
 &= \dynkin(n20\cdots0)t^{n+4} \,.
\end{split}
\end{align}

Adding up all the contributions~(\ref{eq:h2w2})$\sim$(\ref{eq:h2wdel}) and (\ref{eq:h2del2}),
one finds
\begin{align}
 Z_{2,\text{poly}}(t)
&= { -N(t^{-3}-t^{6}) + {(N+2)(N-1)\over2}(t^{-2}-t^{5}) + N(t^{-1}-t^{3}) + {N^2-N+4\over 2}(t^{0}-t^2)
 \over (1-t)^{N} } \nonumber\\
&\qquad + Nt^{-3} + {N^2-N+2 \over 2}t^{-2} + Nt^{-1}\,.
\end{align}
Again, $Z_{2,\text{poly}}(t)$ is non-invariant under the field-antifield symmetry,
but the failure is modest:
\begin{align}
\begin{split}
  Z_{2}(t) &= Z_{2,\text{poly}}(t) - Nt^{-3} - {N^2-N+2 \over 2}t^{-2} - Nt^{-1} \\
\quad\to\quad&  Z_{2}(t) = -(-t)^{2-N}Z_{2}(1/t)\,.
\end{split}
\end{align}
Classically, the elements of the BRST cohomology
that correspond to the missing states are
$b\omega_i$ at $t^{-3}$,
$bJ$ and $bN_{ij}$ at $t^{-2}$,
and $b\del\lambda^{i}$ at $t^{-1}$,
and one can construct the \v{C}ech cocycles
corresponding to those states.

Quantum mechanically, there is a slight
discrepancy in the interpretation
of the symmetric partition function
between the BRST and curved $\beta\gamma$ descriptions.
That is, while both $f_{i}$ and
the \v{C}ech $1$-cocycle corresponding to
$b\del\lambda^{i}$ are in the Hilbert space of
the quantum curved $\beta\gamma$ description,
both are not in the quantum BRST cohomology,
as they form a BRST doublet
(with an exception of the $N=6$ model).
Note, however, that both descriptions still lead to the same
symmetric partition function:
$f_i$ and $b\del \lambda^{i}$ have same charges except for the
statistics so even classically they do not give a net
contribution to the partition function
$\Tr[(-1)^{F}\cdots]$.

\subsection{BRST cohomology and symmetries of partition function}
\label{subsec:BRSTPart}

Since the BRST operator $D$ carries $t$-charge $0$,
the partition function of $D$-cohomology
coincides with that of the unconstrained space of $(\omega,\lambda,b,c)$
in which the cohomology is computed.
This is because the elements not in the cohomology
form BRST doublets and cancel out due to $(-1)^{F}$.
Therefore, the partition function is simply given by~\cite{Grassi:2006wh}
\begin{align}
\label{eq:ZBfull}
Z(q,t) = {1 - t^2 \over (1-t)^{N}}\prod_{h=1}^{\infty}
  {(1-t^2 q^{h})(1-t^{-2}q^{h}) \over (1-tq^h)^{N}(1-t^{-1}q^h)^{N}} \,.
\end{align}
By expanding in $q$, partition functions at fixed Virasoro levels
can be readily obtained.

The full partition function enjoys the following two symmetries,
which turn out to be of fundamental importance.
First is the ``field-antifield symmetry'' we already encountered:
\begin{align}
  Z(q,t) &= -(-t)^{2-N}Z(q,1/t) \,.
\end{align}
As explained above, this symmetry is important to have a nice inner product
after coupling to the fermionic partners $(p_{i},\theta^{i})$.
The other is what we shall call ``$\ast$-conjugation symmetry''
\begin{align}
  Z(q,q/t)  &= -q^{1}t^{-2}Z(q,q/t)\,.
\end{align}
A little computation shows that this symmetry relates the states
at $q^{m}t^{n}$ and those at $q^{1+m+n}t^{-2-n}$,
which suggests the existence of an inner product pairing those.
The inner product is constructed in section~\ref{subsec:innerprod}
using a conjugation operation $\ast$,
which is a generalization of the standard BPZ conjugation~\cite{Belavin:1984vu}.

Although not apparent at this stage,
the inner product responsible for the $\ast$-conjugation symmetry
turns out to be useful for probing the structure of
the BRST cohomology $H^{\ast}(D)$,
because it pairs the states with charges
\begin{align}
  q^{m}t^{n}g^{k}\quad\leftrightarrow\quad q^{1+m+n}t^{-2-n}g^{1-k}\,.
\end{align}
(The exponent of $g$ indicates the ghost number.)
This implies that the elements of $H^{k}(D)$ and $H^{1-k}(D)$
appear in pairs, and we utilize this information to show
that the cohomology is non-vanishing only at ghost numbers $0$ and $1$
(see section~\ref{sec:structBRST}).

Since $H^{0}(D)$ is equivalent to the space of gauge invariant polynomials,
the missing states we found above should be contained in $H^{1}(D)$.
We now explicitly confirm this statement at
several lower mass levels.

\subsection{Extra states in BRST cohomology}
\label{subsec:extraBRST}

In the previous two subsections, we found that the partition function of the BRST cohomology
possesses the field-antifield symmetry while that of the gauge invariant polynomials does not.
We here explicitly construct the elements of the BRST cohomology
and identify the extra states
that are responsible for the discrepancy.

\paragraph{Weight $0$:}

The zero mode contributions to the full partition function~(\ref{eq:ZBfull})
is simply
\begin{align}
  Z_0(t)&= {1 - t^2 \over (1-t)^{N}}\,,
\end{align}
and it coincides with the result obtained from counting
the number of gauge invariant polynomials~(\ref{eq:Z0gi}).
Indeed, since functions of the form $cf(\lambda)$ are never $D$-closed,
and since the functions of the form $(\lambda\lambda)f(\lambda)$ are $D$-exact,
cohomology representatives can be taken as
\begin{align}
  \lambda^{\lpar i_1}\cdots \lambda^{i_n \rpar} \,,
\end{align}
but now with $\lambda$'s unconstrained.
Of course, this is expected from the outset as the BRST construction
is designed to realize what we have just described.

\paragraph{Weight $1$:}

From~(\ref{eq:ZBfull}) one immediately finds
\begin{align}
Z_1(t) &= {-t^{-2} + Nt^{-1} + 1 -t^2 - N t^3 + t^4 \over (1-t)^{N}}\,,
\end{align}
and it possesses the field-antifield symmetry
unlike the level $1$ partition function $Z_{1,\text{poly}}(t)$ of the gauge invariant polynomials.
As expected, $Z_{1}(t)$ contains an extra fermionic state
with respect to $Z_{1,\text{poly}}(t)$:
\begin{align}
 Z_{1}(t)  -  Z_{1,\text{poly}}(t)  = -t^{-2}\,.
\end{align}
Clearly, the cohomology element responsible for $-t^{-2}$ is
the BRST ghost
\begin{align}
  b\,,\quad\text{carrying charges $-q^{1}t^{-2}g^{1}$}\,.
\end{align}
This state is paired with $\mathbf{1}$ at $q^{0}t^{0}g^{0}$ under the $\ast$-conjugation symmetry.
The remaining states correspond to the gauge invariant polynomials~(\ref{eq:h1gi}).
Cohomology representatives basically take the same form,
but for $\omega_{i_1}\lambda^{\lpar i_1}\cdots \lambda^{i_n \rpar}$
it is given by replacing
\begin{align}
  -\omega\lambda \quad\to\quad J_t = -\omega\lambda - 2bc \,.
\end{align}

To summarize, weight $1$ cohomology consists of
$H^{0}(D)|_{h=1}$ (gauge invariant polynomials)
and a single state $b$ from $H^{1}(D)|_{h=1}$.
Note that this is completely consistent
with the structure expected from the $\ast$-conjugation symmetry.
(Gauge invariant states with higher $t$-charges are
paired with states with higher weights
and $\mathbf{1}$ is the only operator which has the partner
in the weight $1$ sector.)

\paragraph{Weight $2$:}

The analysis at weight $2$ is similar.
The partition function respects the field-antifield symmetry and reads
\begin{align}
Z_{2}(t) &= { -N(t^{-3}-t^{6}) + {(N+2)(N-1)\over2}(t^{-2}-t^{5}) + N(t^{-1}-t^{3}) + {N^2-N+4\over 2}(t^{0}-t^2)
 \over (1-t)^{N} }  \,.
\end{align}
The extra states contained are\footnote{%
Here, we removed from $Z_{2,\text{poly}}$
the polynomial $f_{i}$ at $t^{-1}$
since it is not in the quantum BRST cohomology,
as explained above.
Classically, one would add $0=(N-N)t^{-1}$
($f_{i}$ and $b\del\lambda^{i}$) on the right hand side.}
\begin{align}
  Z_{2}(t) - Z_{2,\text{poly}}(t) &= -Nt^{-3} - {N^2-N+2 \over 2}t^{-2} \,,
\end{align}
and one can check that those corresponds to
\begin{align}
  (b\omega_{i},\;bJ,\;bN_{ij})\quad\bigl( \overset{\ast}{\longleftrightarrow}\quad (\lambda^{i},J,N_{ij}) \bigr)\,.
\end{align}
Again, those states all carry ghost number $1$.

\bigskip
At this point, the pattern of the pairing between $H^{0}(D)$ and $H^{1}(D)$
should have become clear.
That is, whenever one has a ghost number $0$ cohomology $F(\omega,\lambda,J,N;\del)$ (gauge invariant polynomial),
the corresponding ghost number $1$ cohomology is obtained
basically by swapping $\omega$ and $\lambda$, and multiplying $b$:
\begin{align}
 \label{eq:H0H1}
  bF(\lambda,\omega,J,N;\del)\quad\overset{\ast}{\longleftrightarrow}\quad F(\omega,\lambda,J,N;\del)\,.
\end{align}
Although the precise representatives for $H^{1}(D)$ in general contain
terms other than $bF$, one can check that the mapping~(\ref{eq:H0H1})
is consistent with the $\ast$-conjugation symmetry.

\subsection{Remark on $\lambda^2=0$ model ($N=1$)}
\label{subsec:remlam2}

Let us make a digression and make a comment on the $N=1$ model.
As mentioned earlier, the constraint for the seemingly simple model
$\lambda^{2}=0$ is in fact reducible
and the use of the naive BRST operator $D=\int b\lambda^{2}$ cannot be justified.
Although $D$ is nilpotent and it makes sense to consider its cohomology,
the cohomology contains unwanted states outside ghost numbers $0$ and $1$.
Let us explicitly identify some unwanted states which are the artifact
of the improper application of the BRST method.

The full partition function of the $D$-cohomology is given by
\begin{align}
\label{eq:charlam2BRST}
Z(q,t) = {1 - t^2 \over (1-t)}\prod_{h=1}^{\infty}
  {(1-t^2 q^{h})(1-t^{-2}q^{h}) \over (1-tq^h)(1-t^{-1}q^h)} \,,
\end{align}
and it possess the two symmetries
\begin{align}
\begin{split}
  Z(q,t)&=t^{1}Z(q,1/t)\,,\quad
  Z(q,t) = -q^{1}t^{-2}Z(q,q/t)\,.
\end{split}
\end{align}
At levels $0$ and $1$, the partition functions read
\begin{align}
\begin{split}
  Z_{0}(t) &= 1+t\,, \\
  Z_{1}(t) &= -t^{-2} + 1 + t - t^{3} \,.
\end{split}
\end{align}
It is easy to obtain the cohomology representatives responsible
for the partition functions.
As usual, $-q^{1}t^{-2}$ corresponds to $b$
and all others but the state at $-q^{1}t^{3}$
correspond to some gauge invariant polynomials.

However, the fermionic state at $-q^{1}t^{3}$
is found to be an unwanted state
\begin{align}
 (-2c\del \lambda + \del c \lambda) \,,
\end{align}
carrying ghost number $-1$.
As can be seen from the naive relation $c\sim \lambda^{2}$,
the occurrence of this state is related to the fact
that the constraint $G\equiv \lambda^{2}=0$ and its derivative
are not independent:
\begin{align}
  2 G \del \lambda = \del G \lambda \,.
\end{align}
(In the standard BRST procedure, one would introduce a pair of
bosonic ghost-for-ghost and extend the BRST operator $D$ to kill this state.)

Finally, let us identify the state paired with $(-2c\del \lambda + \del c \lambda)$
under the $\ast$-conjugation symmetry $q^{m}t^{n}\leftrightarrow q^{m+n+1}t^{-2-n}$.
The conjugate is at $q^{5}t^{-5}$ which is the first term of the
level $5$ partition function
\begin{align}
Z_{5}(t) &= t^{-5} - 3t^{-3} - 5t^{-2} + 7 + 7t - 5t^{3} - 3t^{4}+ t^{6}\,.
\end{align}
The fact that the state at $q^{5}t^{-5}$ is {\em bosonic}
already implies that it is an unwanted state,
since it necessarily carries even ghost number
(which can easily be shown to be non-zero).
The state is
\begin{align}
 b\del b \del \omega\simeq b\del^{2}b \omega\quad(\text{at $q^{5}t^{-5}g^{2}$})
\end{align}
carrying ghost number $2$.
For $N\ge2$ models, one can show that
both $b\del b \del \omega_i$ and $b\del^{2}b \omega_i$ are trivial,
but for $N=1$ (with the ``wrong'' BRST operator) only a linear combination of them is trivial.

\section{Structure of quantum BRST cohomology}
\label{sec:structBRST}

In the previous section, we compared the partition function
of gauge invariant polynomials and that of the BRST cohomology,
and found some extra states in the latter.
This is not strange.
The BRST construction relates the ghost number $0$ cohomology
to the space of gauge invariant polynomials,
but there in general can be cohomologies at non-zero ghost numbers.
In this and the next sections,
we study those extra states in more detail.
First, in this section, we show (for models with $N\ge2$)
that the quantum BRST cohomology is non-vanishing only
at ghost numbers $0$ and $1$,
and that the states in the two sectors come in pairs.
Then in the next section,
we explain how the ghost number $1$ states can be described
in the \v{C}ech or Dolbeault formalisms.

\subsection{Inner product}
\label{subsec:innerprod}

In order to show that the cohomology elements come in pairs,
we first define an inner product
in the space $\calF$ of all operators (not necessarily in the cohomology).
Our inner product is a generalization of the standard BPZ inner product~\cite{Belavin:1984vu},
and it is non-degenerate in the sense
\begin{align}
\forall_{W\in \calF}\; \langle V, W \rangle &=0\quad\to\quad V=0\,.
\end{align}
In other words, every non-zero operator $V$ (not necessarily in the cohomology)
should have at least one operator $W$ satisfying $\langle V,W\rangle\ne0$.

Let us denote the $SL_{2}$ invariant vacuum as
\begin{align}
  \mathbf{1}\;\sim\;\ket|\mathbf{1}>=\ket|0>\,.
\end{align}
In the present case, the vacuum satisfies
\begin{align}
  b_{n}\ket|0>=\omega_{i,n}\ket|0>=0\,,\;(n\ge0)\,,\quad
  c_{n}\ket|0>=\lambda^{i}_{n}\ket|0> = 0\,,\quad(n\ge1)\,,
\end{align}
where as usual the mode expansion of a weight $h$ primary field is
\begin{align}
  \phi(z) &= \sum_{n}\phi_{n}z^{-n-h}\,.
\end{align}
The ``in states'' are constructed by acting the creation operators $(b_{-n-1},c_{-n},\omega_{-n-1},\lambda_{-n})_{n\ge0}$
on the vacuum $\ket|0>$.
Using the state-operator mapping, in states can also be described as
\begin{align}
 \ket|V> = \lim_{z\to0}V(z)\ket|0>\,,
\end{align}
for some operator $V$ which is a polynomial of $b,c,\omega,\lambda$ and their derivatives.

Bosonizing the bosonic $\beta\gamma$ fields as
$(\beta_{i},\gamma_{i})=(\del\xi_{i}\mathe^{-\phi_{i}},\mathe^{\phi_{i}}\eta_{i})$~\cite{Friedan:1985ge}
and setting $\phi=\sum_{i}\phi_{i}$,
the ``out states'' are constructed using
the conjugate operation $\conj$ defined by
\begin{align}
\begin{split}
  \bra<V|&= \ket|V>^{\conj}\quad
\begin{cases} \ket|0>^{\conj} = \bra<\Omega| = \bra<0|\mathe^{-\phi}c_{0}c_{1}\,, \\
  b_{n}^{\conj} = b_{-n-2}\,,\quad
  c_{n}^{\conj} = c_{-n+2}\,,\quad
  \omega_{i,n}^{\conj} = \omega_{i,-n-1}\,,\quad
  {\lambda^{i}_{n}}^{\conj} = \lambda^{i}_{-n+1}\,.
\end{cases}
\end{split}
\end{align}
In terms of conformal fields,
those can be described as a modified BPZ conjugate state
with $\mathe^{-\phi}c\del c$ inserted at infinity:
\begin{align}
  \bra<V|&=\lim_{z\to\infty}\bra<\mathe^{-\phi}c\del c|z^{2L_{0}+J_{0}}V(z)\,.
\end{align}

Now, we define the inner product by the overlap of Fock states
\begin{align}
  \langle V, W \rangle &= \braket<V|W>
\end{align}
with the rule (recall $\bra<\Omega| = \ket|0>^{\ast}$)
\begin{align}
    \bra<\Omega|b_{-1}\ket|0> &= 1\,.
\end{align}
Equivalently, using the notation of conformal field theory,
it can be defined as
\begin{align}
\label{eq:CFTinner}
\begin{split}
 \langle V, W \rangle
&= \lim_{z\to\infty,w\to0}z^{2L_{0}+J_{0}}\llangle V(z)W(w) \rangle \,,
\\
\text{where}\quad& \llangle V(z)W(w) \rangle
= \bra<\mathe^{-\phi}c\del c|V(z)W(w) \ket|0>\,.
\end{split}
\end{align}
Since we inserted $\mathe^{-\phi}c\del c$ at the infinity,
the rule is consistent with the standard rule expected from
anomalies, i.e. $\bra<0|\mathe^{-\phi}c_{0}\ket|0>=1$.

\subsection{Pairing of cohomology}

Up to this point, our argument was general and had nothing to do with
the BRST structure of the system.
We now turn to discuss the implication of the inner product
on the BRST cohomology.
First, since $D(\mathe^{-\phi}c\del c)=0$,
the BRST trivial operators decouple from the inner product~(\ref{eq:CFTinner}).
Therefore,
\begin{align}
  \llangle D(VW) \rangle &= 0\quad\leftrightarrow\quad
  \langle DV, W \rangle + \langle V,DW\rangle = 0\,.
\end{align}
Using this property, it is easy to show that the cohomology elements come in pairs.

Let us split the space of operators $\calF$ as follows:
\begin{align}
\begin{split}
  \calF&=\calA+\calB+\calH
 = \begin{cases}
  \calA\colon&\text{$D$-non-closed\,,} \\
  \calB\colon&\text{$D$-exact\,,} \\
  \calH\colon &\text{$D$-cohomology\,.}
\end{cases}
\end{split}
\end{align}
Although there is no canonical way to achieve the splitting between $\calB$ and $\calH$,
one can argue that the inner product~(\ref{eq:CFTinner})
induces a non-degenerate inner product on the cohomology $\calH$.
This follows from the following two properties:
\begin{enumerate}
\item $V\in \calB$ and $\langle V,W\rangle\ne0$ $\to$ $W\in \calA$ ($DW\ne0$)
\item $V\in \calA$ $\to$ $\exists W\in\calB$ s.t. $\langle V,W\rangle \ne0$
\end{enumerate}

\paragraph{Proof of 1.}

Let $V_c$ denote a conjugate of $V\in \calF$, i.e. $\langle V, V_c \rangle \ne0$.
(It is not unique but we do not rely on the uniqueness of $V_c$
in the following arguments.)
Since $V$ is $D$-exact, it can be written as $V=DU$ for some $U$.
For all $V_c$, one has
\begin{align}
  0 &= \llangle D(UV_c) \rangle = \llangle (DU)V_c \rangle + \llangle U (DV_c) \rangle\,,
\end{align}
but since $\llangle(DU)V_c\rangle=\llangle VV_c\rangle \ne0$, it follows that
$\llangle U (DV_c) \rangle\ne0$ which in turn implies $DV_c\ne0$
(and $U_c = DV_c$).

\paragraph{Proof of 2.}

Denote $U\equiv DV\ne0$ and let $U_c$ be one of its conjugate.
Then,
\begin{align}
  0 &= \llangle D(U_c V) \rangle = \llangle (DU_c) V \rangle + \llangle U_c (DV) \rangle\,,
\end{align}
and since $\llangle U_c (DV) \rangle=\llangle U_c U \rangle \ne0$,
one finds $V_c=DU_c$.

\bigskip
Now, the property 1 implies $\langle\calB,\calB\rangle=\langle\calB,\calH\rangle=0$,
while the property 2 implies that the matrix $\langle\calA,\calB\rangle$ has the maximal rank.
Thus, schematically, the inner product for the full space looks like the
first matrix in the diagram below.
(The star $\star$ signifies the maximal rank
and the question mark $?$ designates blocks whose properties are
unknown.)
This then implies that one can choose appropriate representatives for the cohomology $\calH$
so that $\langle\calA,\calH\rangle=0$ (the second matrix).
Finally, the non-degeneracy of the submatrix $\langle\calH,\calH\rangle$ follows
from that of the full matrix.
\begin{align*}
  \bordermatrix{ & \calA & \calB & \calH \cr
  \calA & ? & \star & ? \cr
  \calB & \star & 0 & 0 \cr
  \calH & ? & 0 & ? \cr}
\to   \bordermatrix{ & \calA & \calB & \calH \cr
  \calA & ? & \star & 0 \cr
  \calB & \star & 0 & 0 \cr
  \calH & 0 & 0 & ? \cr}
\to   \bordermatrix{ & \calA & \calB & \calH \cr
  \calA & ? & \star & 0 \cr
  \calB & \star & 0 & 0 \cr
  \calH & 0 & 0 & 1 \cr}\quad(\because\det\ne0) \,.
\end{align*}

\subsection{Vanishing theorem for $H^{k}(D)$ with $k\ne0,1$}

Using the pairing of cohomologies just described,
one can show that the BRST cohomology is non-vanishing
only at ghost numbers $0$ and $1$.
To see this, recall that the quantum charges of a state and
its $\ast$-conjugate are related as
\begin{align}
  q^{m}t^{n}g^{k}\quad\leftrightarrow\quad q^{m+n+1}t^{-2-n}g^{1-k} \,,
\end{align}
where $m$ is the weight, $n$ is the $t$-charge,
and $k$ is the ghost number.
Our claim is then equivalent to the assertion $H^{k}(D)=0$ ($k<0$).
That is, there are no cohomology elements with negative ghost numbers
(which means the number of $c$ ghosts is
strictly greater than that of $b$ ghosts).
$H^{k}(D)=0$ ($k<0$) is true more or less by construction,
but let us briefly sketch why it is the case.

In the BRST formalism,
$c$-type ghosts represent the constraint ($c\overset{D}{\to}\lambda\lambda$)
and the formalism is designed so that the $c$-type ghosts
do not contribute to the cohomology in any important way.
By construction,
there are no negative ghost number cohomologies without $b$;
whenever there is a $D$-closed operator of the form
\begin{align}
 f_{k}(\omega,\lambda,c;\del) &= \sum_{\{N\}} \del^{N_1}c\cdots \del^{N_k}c\; f_{N_1 \cdots N_k}(\omega,\lambda;\del)\,,
\end{align}
one can show that it is $D$-exact.
(If this is not the case, additional $c$-type ghosts must be introduced
and the BRST charge must be extended to make it $D$-exact,
$c'\overset{D}{\to}f_k$.
This will be the case when the constraints are reducible.)
In fact, it can be shown that the same is true for
the negative ghost number operators
with both $b$ and $c$~\cite{Henneaux:1992ig},
\begin{align}
 f_{k}(\omega,\lambda,b,c;\del) &=
  \sum_{i\ge0}\sum_{\{M,N\}} \del^{M_1}b\cdots \del^{M_{i}}b\;
    \del^{N_1}c\cdots \del^{N_{k+i}}c \;
  f_{M_1\cdots M_{i} N_1 \cdots N_{k+i}}(\omega,\lambda;\del)\,.
\end{align}
If $f_{k}$ is $D$-closed, the terms without $b$ ($i=0$)
can be written in a $D$-exact form,
modulo terms with at least one $b$ ($i\ge1$).
After subtracting the $D$-exact piece just mentioned,
the equation $Df_{k}=0$ implies that the coefficients of $\del^{M_1}b$,
i.e. $\del^{N_1}c\cdots \del^{N_{k+1}}c\; f_{M_{1}N_{1}\cdots N_{k+1}}$,
are $D$-closed (and hence $D$-exact) modulo terms with at least two $b$'s.
Therefore,
$f_{k}$ is $D$-exact modulo terms with at least two $b$'s ($i\ge2$).
Proceeding inductively in number of $b$'s,
one can show that $f_{k}$ is $D$-exact.

Therefore, we conclude that $H^{k}(D)=0$ ($k<0$),
and hence $H^{k}(D)=0$ ($k>1$)
via the $\ast$-conjugation symmetry.

\section{Relating BRST, \v{C}ech and Dolbeault cohomologies}
\label{sec:equivcohoms}

In the previous section, we found that the BRST cohomology includes
extra states at ghost number $1$
that do not correspond to gauge invariant polynomials.
Those states were important for having the field-antifield symmetry.
We here sketch the equivalence between the BRST and \v{C}ech/Dolbeault
descriptions, by giving a mapping that relates the classical pieces
of the cohomology element.
In particular we shall explain how the ghost number $1$ extra states
are described in the intrinsic \v{C}ech/Dolbeault framework.

Since the BRST and the intrinsic curved $\beta\gamma$ frameworks
use different normal ordering prescriptions,
the quantum BRST and \v{C}ech-Dolbeault cohomologies
differ in general.
This indeed happens for our models.
However, as we have mentioned several times,
our partition function $\Tr[(-1)^{F}\cdots]$ is insensitive to
such discrepancies.

\subsection{BRST, \v{C}ech and Dolbeault cohomologies}
\label{subsec:4cohoms}

It is convenient to introduce the following four cohomologies,
which classically give different representation of a same space:
\begin{enumerate}
\item Minimal BRST: Cohomology of $D$
\item Non-minimal BRST: Cohomology of $D+\deldol$
\item Dolbeault cohomology $\deldol$ (of gauge invariant operators)
\item \v{C}ech cohomology (of gauge invariant operators)
\end{enumerate}
As explained in~\ref{subsec:modelsBC},
the notion of ``gauge invariance'' in curved $\beta\gamma$ frameworks
(for $N\ge4$ models)
is a simple way to refer to the operator intrinsic
to the target space $X$
but by using the extrinsic coordinate $(\omega,\lambda)$ of the space where $X$ is embedded.
We find it especially useful when comparing to the BRST framework.

Although we already described most of them,
let us recapture the definitions of each.

\paragraph{Minimal BRST cohomology}

This is simply the standard BRST cohomology of $D=\int b(\lambda\lambda)$,
computed in the space of polynomials of unconstrained $(\omega,\lambda)$,
BRST ghosts and their derivatives,
\begin{align}
  f(\omega,\lambda,b,c;\del)\,.
\end{align}
By construction, the ghost number $0$ cohomology $H^{0}(D)$
is isomorphic to the space of
gauge invariant polynomials of the constrained system.
On the other hand, as we observed above,
there are also the operators with non-zero ghost numbers,
but the higher cohomology is non-empty only at ghost number $1$
(where $b$ carries ghost number $+1$).
Obtaining the expressions for those extra states
in the curved $\beta\gamma$ framework, i.e. in the \v{C}ech/Dolbeault cohomologies,
is the goal of the present section.

\paragraph{Non-minimal BRST cohomology}

Closely related to the minimal BRST cohomology
is what we call non-minimal BRST cohomology.
This is defined by introducing the
{\em unconstrained} non-minimal variables
$(\overline{\omega}^{i},\overline{\lambda}_{i};s^{i},r_{i})$
and extending the BRST operator as
\begin{align}
 \overline{D} &= D + \deldol \,,\quad
\deldol = -r_{i}\overline{\omega}^{i} \sim \targetd\overline{\lambda}_{i}{\del \over \del \overline{\lambda}_{i}}\,.
\end{align}
The cohomology of $\overline{D}$ is computed in the space of functions of the form
\begin{align}
  \label{eq:nonminimalspace}
  f(\omega,\lambda,\overline{\omega},\overline{\lambda},r,s,b,c;\del)\,,
\end{align}
where now $f$ can diverge as fast as $(\lambda\overline{\lambda})^{-n}$ for $n<N$.

The restriction on the order of poles is important.
If one allows the functions that diverge as fast as $(\lambda\overline{\lambda})^{-N}$,
there will be extra cohomology elements due to the operator
\begin{align}
{\overline{\lambda}_{[i_{1}}r_{i_{2}}\cdots r_{i_{N}]} \over (\lambda\overline{\lambda})^{N} } \,,
\end{align}
which do not have counterparts in minimal BRST cohomology.

We introduced the non-minimal variables as unconstrained variables,
however, it should be noted that
they do not affect the cohomology
even if they are considered to be constrained,
as long as the minimal variables are {\em unconstrained}.
Whether constrained or not,
the non-minimal variables can appear only in the combinations
$\overline{\lambda}_{i}(\lambda\overline{\lambda})^{-1}$ and $r_{i}(\lambda\overline{\lambda})^{-1}$
(other combinations of non-minimal variables are irrelevant
due to the usual quartet mechanism),
and one can switch between the two viewpoints
by simply forgetting/imposing the non-minimal constraint.

Non-minimal BRST description is a hybrid between minimal BRST and Dolbeault
languages, and provides the key to relate the minimal BRST and Dolbeault
descriptions.  The space on which $\overline{D}$ acts~(\ref{eq:nonminimalspace})
is doubly graded by the BRST ghost number
and the Dolbeault form degree.

\paragraph{Dolbeault cohomology}

We now turn to the description of cohomologies in the curved $\beta\gamma$ schemes.
The cohomology of the differential operator $\deldol=-r_{i}\overline{\omega}^{i}$
is computed in the space of functions of the form
\begin{align}
  f(\omega,\lambda,\overline{\omega},\overline{\lambda},r,s;\del)\,.
\end{align}
Again, $f$ is allowed to diverge as $(\lambda\overline{\lambda})^{-n}$ ($n<N$),
but additionally it must be gauge invariant
(if one is to write $f$ using the extrinsic coordinates $(\omega,\lambda)$).

The cohomology splits naturally into two families.
One family is the globally defined gauge invariant polynomials
without poles in $(\lambda\overline{\lambda})$.
The other corresponds to the operators in the higher BRST cohomology.
The BRST ghost number corresponds to the form degree of the Dolbeault cohomology,
i.e. the number of $r_{i}$'s
(that can only appear in the combination $(\lambda\overline{\lambda})^{-1}r_{i}$).
Since operators diverging too fast as $(\lambda\overline{\lambda})\to0$
are troublesome for the computation of amplitudes~\cite{Berkovits:2006vi},
we do not want to have cohomologies at too high degrees.

\paragraph{\v{C}ech cohomology}

Finally, the \v{C}ech-type description of the cohomology is obtained from
the Dolbeault description using the usual \v{C}ech-Dolbeault correspondence.
Elements of the cohomology will be the \v{C}ech $n$-cocycles of the form
\begin{align}
f &=   (f^{A_{0}\cdots A_{n}})
 = f^{A_{0}\cdots A_{n}}(\omega,\lambda;\del)\,,\quad (n\ge0)\,,
\end{align}
where $f^{A_0\cdots A_{n}}$ denotes a collection of gauge invariant functions
defined on overlaps $U_{A_0\cdots A_n}=U_{A_0}\cap\cdots \cap U_{A_n}$.
On $U_{A_0\cdots A_n}$, $f$ is allowed to have poles in $\lambda^{A_{i}}$ ($i=0\sim n$).
The degrees of cochains are related to the form degree in Dolbeault description,
and hence to the BRST ghost numbers.
The gauge invariant polynomials are represented as $0$-cocycles,
and the extra states at ghost number $n$ are
represented as $n$-cocycles that are defined modulo $n$-coboundaries.

\subsection{Classical equivalence of various cohomologies}

Operators in the four cohomologies in the previous subsection
can be related as indicated in the following figure.
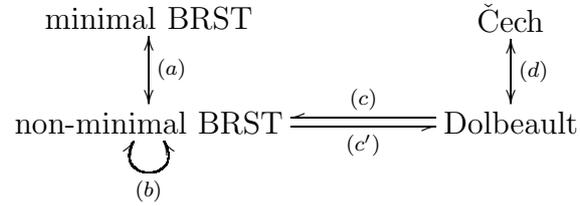
\begin{align*}
\xymatrix{
{ \txt{minimal BRST}\ar@{<->}[d]^-{(a)}  } && { \txt{\v{C}ech}\ar@{<->}[d]^-{(d)} }\\
{ \txt{non-minimal BRST}\ar@^{<-}@<.3ex>[rr]^-{(c)}\ar@_{->}@<-.3ex>[rr]_-{(c')}\ar@{<->}@(dl,dr)_-{(b)}  } && {\txt{Dolbeault}} \\
  &&
}
\end{align*}
\begin{description}
\item[\quad$(a)$] Adding/removing non-minimal quartet under $\deldol=-r\overline{\omega}$
\item[\quad$(b)$] Different choice of cohomology representatives
\item[\quad$(c)$] Embedding to ``extrinsic'' space of free fields
\item[\quad$(c')$] Restriction to ``intrinsic'' (or gauge invariant) operators on $X$
\item[\quad$(d)$] Standard \v{C}ech-Dolbeault mapping (partition of unity)
\end{description}
The idea here is to use the non-minimal BRST cohomology $H^{\ast}(D+\deldol)$
to bridge between the BRST and curved $\beta\gamma$ schemes,
as the following figure indicates:

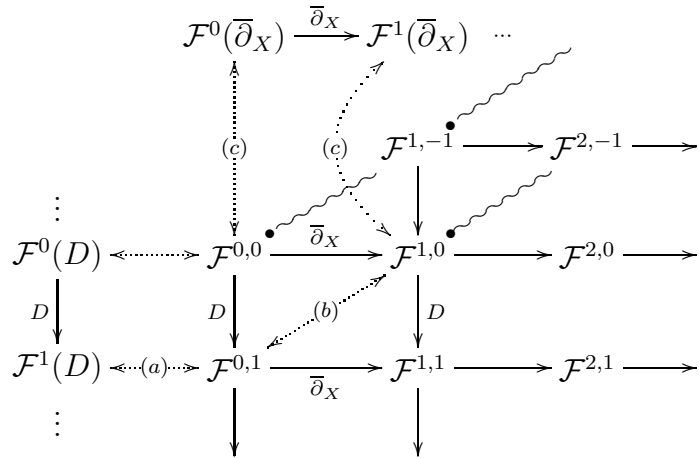
\begin{align*}
\xymatrix{
  &  \calF^{0}(\deldol)\ar@{->}[r]^{\deldol}  \ar@{<.>}[dd]|-{(c)}
    &  \calF^{1}(\deldol)\ar@{}[r]|{\cdots} \ar@{<.>}@/_6ex/[dd]|-{(c)}
      & & \\
  & & \calF^{1,-1}\ar@{->}[d] \ar@{->}[r] \ar@{{*}~}[ur]
      & \calF^{2,-1}\ar@{->}[r] &\\
\calF^{0}(D)\ar@{<.>}[r]\ar@{->}[d]_{D}\ar@{}[u]|{\vdots}
  & \calF^{0,0}\ar@{->}[d]_{D}\ar@{->}[r]^{\deldol}\ar@{{*}~}[ur]
    &\calF^{1,0}\ar@{->}[d]^{D}\ar@{->}[r]  \ar@{{*}~}[ur]
      & \calF^{2,0}\ar@{->}[r]& \\
\calF^{1}(D)\ar@{<.>}[r]|-{(a)}\ar@{}[d]|{\vdots}
  & \calF^{0,1}\ar@{->}[d]\ar@{->}[r]_{\deldol} \ar@{<.>}[ur]|-{(b)}
    &\calF^{1,1}\ar@{->}[d]\ar@{->}[r]
      & \calF^{2,1}\ar@{->}[r] & \\
  &&&&
}
\end{align*}
\gdef\FigEmbNM{1}
\centerline{Figure \arabic{section}.\FigEmbNM: Embedding to the non-minimal BRST cohomology}

\bigskip
In the figure, we put the minimal $D$-cohomology on the left-most column
and the $\deldol$-cohomology on the top row.
The non-minimal BRST cohomology of $(D+\deldol)$ is
graded by the sum of BRST ghost number and the Dolbeault form degree
(number of $r$'s), which runs diagonally from north-west to south-east.

Both $D$ and $\deldol$ cohomologies can be embedded
in the $(D+\deldol)$-cohomology
as indicated by the arrows $(a)$ and $(c)$.
A ghost number $k$ element of the $D$-cohomology
can be regarded as a $(D+\deldol)$-cohomology element
with degree $(0,k)$.
A degree $n$ element of the $\deldol$-cohomology
can also be regarded as a $(D+\deldol)$-cohomology element,
but this time the corresponding element in general
has multiple (bi)degrees $\sum_{k\ge0}\calF^{n+k,-k}$.

Once the embedding into the non-minimal $(D+\deldol)$-cohomology is achieved,
the cohomologies of $D$ and $\deldol$ simply
correspond to different choices of cohomology representatives,
where the non-minimal variables are absent (minimal BRST),
and the ($b$-type) BRST ghosts are absent (Dolbeault),
as indicated by the arrow $(b)$.

\subsubsection{Embedding to non-minimal BRST cohomology}

\paragraph{Embedding $(a)$}

First, let us describe the embedding
of the minimal BRST cohomology to the non-minimal BRST cohomology.
Since $D$ and $\deldol$ anticommute, cohomology of $\overline{D}$ is the
cohomology of $D$ computed in the cohomology of $\deldol$.
Note that the $\deldol$-cohomology here is computed in the space
where the constraint for the minimal variable $\lambda$ is absent.
Hence, provided one restricts the order of poles in $(\lambda\overline{\lambda})$,
the cohomology of $\deldol$ is simply the space without non-minimal variables.
That is, all elements of the $\deldol$-cohomology have
representatives of the form
\begin{align}
  f(\omega,\lambda,b,c;\del)\quad{\text{\small (no poles in $\lambda$)}}\,,
\end{align}
which is nothing but the space where the minimal BRST cohomology
is computed.

\paragraph{Embedding $(c)$}

For the models at hand, a Dolbeault cohomology element
with form degree $n$ can be represented
by a gauge invariant function $f^{n}$.
Classically, from $f^{n}$, one gets an operator $f^{n,0}$ living in the space $\calF^{n,0}$,
by simply forgetting the constraint $(\lambda\lambda)=0$.
In contrast to the elements of the minimal BRST cohomologies above, however,
$f^{n,0}$ is not necessarily $(D+\deldol)$-closed.
Nevertheless, following the standard argument in the BRST formalism,
$f^{n,0}$ can be extended to
the form $\Hat{f}^{n}= \sum_{k\ge0}\Hat{f}^{n+k,-k}$
so that
\begin{align}
\label{eq:nonminchain}
(D+\deldol)\Hat{f}=0\quad\Leftrightarrow\quad
\left\{
\begin{array}{rcl}
  D \Hat{f}^{n,0}&=&0\,,  \\
  D \Hat{f}^{n+1,-1} + \deldol \Hat{f}^{n,0}&=&0 \,,\\
    & \vdots& \\
  D \Hat{f}^{n+p-1,-p+1} + \deldol \Hat{f}^{n+p-2,-p+2}&=&0 \,,\\
  \deldol \Hat{f}^{n+p,-p}&=&0 \,,
\end{array}
\right.
\end{align}
for some $p$, or, more pictorially,
\begin{align*}
 \xymatrix@!=3ex{
&  \Hat{f}^{n,0} \ar@{->}[ld]|{D} \ar@{->}[rd]|{\deldol} & { + }
&  \Hat{f}^{n+1,-1} \ar@{->}[ld]|{D} \ar@{->}[rd]|{\deldol} & { +\;\;\cdots } & { + }
&  \Hat{f}^{n+p,-p} \ar@{->}[ld]|{D} \ar@{->}[rd]|{\deldol}  \\
  0 &  & 0 &  & 0 & 0 &  & 0}
\end{align*}
That is, a Dolbeault cohomology element with degree $n$
corresponds to a sequence of
non-minimal operators with its ``head'' in $\calF^{n,0}$
(see figure~\arabic{section}.\FigEmbNM).

For completeness, let us briefly sketch the procedure to obtain the
sequence $\Hat{f}^{n}=\sum_{k\ge0}\Hat{f}^{n+k,-k}$,
starting from a constrained operator $f^{n}$.
Firstly, the unconstrained operator $f^{n,0}$ naively obtained from $f^{n}$
is not necessarily $D$-closed,
but it satisfies
\begin{align}
\label{eq:Dfgh}
\begin{split}
  D f^{n,0} &= g^{n,1} \approx 0 \quad\text{(gauge invariance of $f^{n}$)}\,, \\
  \deldol f^{n,0} &= g^{n+1,0} \approx0 \quad \text{($\deldol$-closed condition of $f^{n}$)}\,,
\end{split}
\end{align}
for some $g^{n,1}\in\calF^{n,1}$ and $g^{n+1,0}\in\calF^{n+1,0}$.
As indicated in the first formula,
gauge invariance of the original $f^{n}$ implies
that $g^{n,1}$ vanishes on $(\lambda\lambda)=0$,
and of course $g^{n,1}$ contains one $b$.
Hence, $g^{n,1}$ can be written
as $D\Bar{f}^{n,0}$ where $\Bar{f}^{n,0}$ is different from $f^{n,0}$.
For example, for $f^{n,0}=\lambda\omega$, one has $\deldol f^{0,0}=0$ and
\begin{align}
\label{eq:f00}
  D f^{0,0} &=  g^{0,1} = 2b(\lambda\lambda) = D \Bar{f}^{0,0}\quad
\text{where}\quad \Bar{f}^{0,0} = {(\overline{\lambda}\omega)(\lambda\lambda) \over \lambda\overline{\lambda} } \,.
\end{align}
By setting $\Hat{f}^{n,0}=f^{n,0}-\Bar{f}^{n,0}$,
one obtains the ``head'' of the chain $\Hat{f}^{n}$ in~(\ref{eq:nonminchain}).

On the other hand, using $\{\deldol\,, D\} = D^{2}=0$
and the second equation in~(\ref{eq:Dfgh}),
one finds after a little computation that
\begin{align}
\deldol \Hat{f}^{n,0} &= \Hat{g}^{n+1} \; (\equiv g^{n+1,0} - \Bar{g}^{n+1,0})\,,
\end{align}
where $g^{n+1,0}= \deldol f^{n+1,-1}$ and $\Bar{g}^{n+1,0}=\deldol\Bar{f}^{n,0}$ are
separately $D$-closed.
In fact both are weakly zero and hence are $D$-exact.
For example, $\Bar{f}^{0,0}$ in~(\ref{eq:f00}) satisfies
\begin{align}
\label{eq:JembDoltoBRST}
  \deldol\Bar{f}^{0,0}
&= D\Bar{f}^{1,-1}\quad\text{where}\quad
\Bar{f}^{1,-1} = \Bigl(c {(\lambda\overline{\lambda})(rw) - (\overline{\lambda}\omega)(\lambda r) \over (\lambda\overline{\lambda})^{2}}\Bigr) \,.
\end{align}
($\deldol f^{0,0}=0$ in this case.)
Choosing an operator $\Hat{f}^{n+1,-1}$ satisfying $D\Hat{f}^{n+1,-1}=g^{n+1,0}$,
the sum $\Hat{f}^{n,0}+\Hat{f}^{n+1,-1}$ solves
the master equation~(\ref{eq:nonminchain}) to the second line.

Proceeding in a similar manner,
one can iteratively determine $\Hat{f}^{n+k,-k}$ ($k>1$)
as follows
\begin{align}
  \Hat{g}^{n+k+1,-k} &\equiv \deldol\Hat{f}^{n+k,-k}
\;\to\;   D\Hat{g}^{n+k+1,-k} = 0\;\to\;\Hat{g}^{n+k+1,-k} = D\Hat{f}^{n+k+1,-k-1}\,.
\end{align}
In general, $D^{2}=\{D\,,\deldol\}=0$ implies that $\Hat{g}^{n+k+1,-k}$
defined by the first equation is $D$-closed.
Then since $D$ has no cohomologies at negative degrees,
$\Hat{g}^{n+k+1,-k}$ is found to be $D$-exact.

\subsubsection{Various descriptions of the $b$ ghost}
\label{sec:b}

Before explaining the general relation between BRST and Dolbeault descriptions
embedded in the non-minimal BRST cohomology,
let us study how the ghost $b$ is described in various cohomologies.
Since the quantum BRST cohomology $H^{k}(D)$ is non-vanishing
only at ghost numbers $0$ and $1$,
clearly the ghost $b$ (which is the lowest mass operator in $H^{1}(D)$)
plays a special role among others.

\paragraph{Dolbeault description}

As explained above,
$b \in H^{1}(D)$ is also in the cohomology of the non-minimal BRST operator $\overline{D}=D+\deldol$.
But since inverse powers of $\lambda\overline{\lambda}$ can be used in
the non-minimal formulation,
operators can have drastically different expressions in this cohomology.
Indeed, using the relation
\begin{align}
 \label{eq:bDsome}
  b &= D\biggl( {\overline{\lambda}\omega \over 2\lambda\overline{\lambda} } \biggr) \,,
\end{align}
one can represent $b$ in a gauge where all BRST ghosts are absent:
\begin{align}
\begin{split}
  b &\simeq -\deldol \biggl( {\overline{\lambda}\omega \over 2\lambda\overline{\lambda} } \biggr)  \\
    &= {  (\lambda r)(\overline{\lambda}\omega) - (\lambda\overline{\lambda})(r  \omega) \over 2(\lambda\overline{\lambda})^{2} }    \,.
  \end{split}
\end{align}
Since there are no ghosts in the final expression,
it is easy to identify the corresponding operator in the Dolbeault cohomology:
\begin{align}
\label{eq:bDol}
  \Bar{b} &= { (\lambda r)(\overline{\lambda}\omega) - (\lambda\overline{\lambda})(r  \omega) \over 2(\lambda\overline{\lambda})^{2} }  \,.
\end{align}
While $\Bar{b}$ is trivially $\deldol$-closed
(as it is formally a $\deldol$ of a gauge {\em non}-invariant quantity),
it is not a $\deldol$ of a gauge invariant operator
and hence is in the Dolbeault cohomology.

Although $b$ and $\Bar{b}$ look identical,
we emphasize that they are conceptually quite different.
In particular, in the space where $\Bar{b}$ is defined,
the constraint $(\lambda\lambda)=0$ and the associated gauge invariance are in effect,
while they are not for the space where $b$ is defined.

Quantum mechanically,
depending on the normal ordering prescription used to define $\Bar{b}$,
there can be quantum improvement terms of the form
$(\lambda\overline{\lambda})^{-2}(\overline{\lambda}\del r - r\del\overline{\lambda})$
to assure that $b$ is $\deldol$-closed.

\paragraph{\v{C}ech description}

As usual, the \v{C}ech and Dolbeault cohomologies are related by
the partition of unity on the target space $X$~\cite{Witten:2005px}.
As described in appendix~\ref{app:geomcone}, $X$ can be covered using
$2N$ patches $U_A$ ($A=1\sim 2N$),
where in $U_{A}$ a certain component of $\lambda$ which we denote $\lambda^A$ is non-vanishing.
The partition of unity and an associated differential is given by
\begin{align}
\begin{split}
  \rho_{A} &= {\overline{\lambda}_{A}\lambda^{A} \over \lambda\overline{\lambda}}\,,\quad
  \sum_{A}\rho_{A} = 1
\\
  \targetdbar\rho_{A} &= {(\lambda\overline{\lambda})r_{A}\lambda^{A} - (\lambda r)\overline{\lambda}_{A}\lambda^{A}\over (\lambda\overline{\lambda})^{2}} \,.
\end{split}
\end{align}
(Here and hereafter, we do not use the Einstein summation convention for the index $A$.)
Now, the state $\Bar{b}$~(\ref{eq:bDol}) is written as
\begin{align}
  \Bar{b} &= -\sum_{A,B}{\lambda^{[A}\omega^{B]} \over \lambda^{A}\lambda^{B} }\rho_A\targetdbar \rho_B \,,
\end{align}
and hence it corresponds to a \v{C}ech $1$-cochain
\begin{align}
  \Check{b} &= (b^{AB}) = -{ 2\lambda^{[A}\omega^{B]} \over \lambda^{A}\lambda^{B} }  \,.
\end{align}
While $\Check{b}$ trivially satisfies the cocycle condition
as it is formally a $\delcech$ (difference) of two gauge {\em non}-invariant $0$-cochains,
\begin{align}
  \Check{b} &= \delcech\Bigl( {\omega^{A} \over 2\lambda^{A}} \Bigr) = {\omega^{A} \over 2\lambda^{A}}-{\omega^{B} \over 2\lambda^{B}}\,,
\end{align}
it is not a difference of gauge invariant $0$-cochains
and hence is in the \v{C}ech cohomology.
Of course, this corresponds to the fact that
$\Bar{b}$ is a $\deldol$ of gauge {\em non}-invariant function
but not a $\deldol$ of gauge invariant function.
Using the local coordinates on the overlaps $U_{A}\cap U_{B}$,
it can be written as
\begin{align}
\begin{split}
  \Check{b} &= (b^{AB}) = (b^{1,\Tilde{1}}, b^{1,2},\cdots) \,,\\
\text{where}\quad&  b^{1,\Tilde{1}} = {\varrho -{1\over2}(u\cdot u)(u\cdot \del u)\over g^2(u\cdot u)}
  = {\Tilde{\varrho} -{1\over2}(\Tilde{u}\cdot \Tilde{u})(\Tilde{u}\cdot \del\Tilde{u})\over \Tilde{g}^2(\Tilde{u}\cdot\Tilde{u})}
\,,\quad b^{1,2}=\cdots \,.
\end{split}
\end{align}

\subsubsection{Classical mapping between BRST and Dolbeault descriptions and quantum discrepancy}

It is straightforward to extend the mapping for the $b$ ghost above
to other operators in the cohomology.
For the operators in $H^{0}(D)$ (those corresponding to usual gauge invariant
polynomials), the mapping in essence
is simply a matter of dropping and recovering the
ghost contribution in the $t$-charge current\footnote{%
Note that BRST ghosts are rotation singlet in our models,
so the current $N_{ij}$ does not contain the ghosts.}:
\begin{align}
  J &= -\omega\lambda - 2bc \quad \leftrightarrow\quad J = -\omega\lambda \,.
\end{align}

As for the operators in $H^{1}(D)$,
the mapping works just as in the case of $b$ ghost.
One simply gets rid of the $b$ (or its derivative)
by using the relation~(\ref{eq:bDsome});
this leads to the expression of the non-minimal cohomology element
in a gauge where the $b$ ghost is absent
(apart from those contained in $J$'s).

Classically,
the higher cohomologies $H^{k}(D)$ ($k>1$) are not empty
as opposed to the quantum case.
For example, a pair of operators $b\del b$ and $b(\omega\omega)$
with charges $q^{3}t^{-4}$ are both in the classical cohomology.
(Quantum mechanically, those form a BRST doublet.)
Using the fact that $\del^{n}b=D\del^{n}(\overline{\lambda}\omega/2\lambda\overline{\lambda})$
and $[\deldol,D]=0$, however,
one can map those higher cohomology elements
into the non-minimal gauge by eliminating
one unit of ghost charge at a time.

Quantum mechanically, a pair $(\Hat{f},\Hat{g})$ of the elements of classical $(D+\deldol)$-cohomology
may drop out from the cohomology by forming a doublet $D+\deldol\colon \Hat{f}\to \Hat{g}$.
Since the curved $\beta\gamma$ and BRST descriptions use different
normal ordering prescriptions,
it is not assured that this happens
if and only if the corresponding elements
in the Dolbeault cohomology form a doublet as $\deldol\colon f\to g$.
Indeed, there are mismatches between the two descriptions
as explained at the end of section~\ref{subsec:GIpoly}.

\subsubsection{Examples of the mapping}

Now, let us illustrate the mapping by translating some
specific operators from BRST to \v{C}ech-Dolbeault languages.

\paragraph{Example: $t$-charge current $J$}

First, consider the $t$-charge current $J=-\omega_{i}\lambda^{i}-2bc$.
From
\begin{align}
\begin{split}
 D\biggl( {(\overline{\lambda}\omega)c \over \lambda\overline{\lambda} } \biggr)
 &= 2bc + {(\overline{\lambda}\omega)(\lambda\lambda)  \over \lambda\overline{\lambda} }+ {2(\del \lambda\overline{\lambda}) \over \lambda\overline{\lambda}}\,, \\
 \deldol \biggl( {(\overline{\lambda}\omega) c\over \lambda\overline{\lambda} }\biggr)
 &= { (r\omega) c\over \lambda\overline{\lambda} }  - { (\lambda r)(\overline{\lambda}\omega) c\over (\lambda\overline{\lambda})^{2} }\,,
\end{split}
\end{align}
one finds the following representation of $J$ in the non-minimal BRST cohomology:
\begin{align}
\label{eq:JmappedfromBRST}
\begin{split}
  J &\simeq
    -\omega\lambda + {2(\del \lambda\overline{\lambda}) \over \lambda\overline{\lambda}}  \\
 &\qquad + {(\overline{\lambda}\omega)(\lambda\lambda)  \over \lambda\overline{\lambda} }
  + c { (\lambda\overline{\lambda})(r\omega)  -  (\lambda r)(\overline{\lambda}\omega) \over (\lambda\overline{\lambda})^{2} }\,.
\end{split}
\end{align}
Apart from the second term, which is a quantum correction,
the expression of $J$ is precisely the one we obtained
in~(\ref{eq:JembDoltoBRST})
by embedding the Dolbeault cohomology to the non-minimal BRST cohomology.

The normal ordering in~(\ref{eq:JmappedfromBRST})
is that of the free fields.
Since $c=(\lambda\lambda)=0$ in the Dolbeault language,
the $t$-charge current should look like
\begin{align}
  \Bar{J}
 &= -\omega\lambda + {2(\del \lambda\overline{\lambda}) \over \lambda\overline{\lambda}} \,,
\end{align}
where $\omega$, $\lambda$ and $\overline{\lambda}$ are parameterized by
some independent variables.
The second term represents some quantum correction,
but as $\omega$ and $\lambda$ are no longer free,
there seems to be no reason to believe the value of its coefficient.
We, however, observe that this value can be understood
intuitively as the anomaly
coming from the constraint per se.

Note that in a local coordinate one classically has
\begin{align}
  \Bar{J} &= -\varrho \sim \omega\lambda\,,
\end{align}
where $\varrho$ is the conjugate to $\varphi$
parameterizing the length of $\lambda$ (see appendix~\ref{app:geomcone}).
Quantum mechanically, {\em if $\omega$ and $\lambda$ were free fields}, this is modified to
\begin{align}
  \Bar{J} &= -\varrho - {N\over2}\del\varphi\quad\Bigl(\to\quad \Bar{J}(z)\Bar{J}(w) = {-N \over (z-w)^{2}}\Bigr)\,,
\end{align}
receiving the correction from the usual free field chiral anomalies.
However, some units of the background charge are absent due to constraint,
and this is exactly represented by $2(\del\lambda\overline{\lambda})/(\lambda\overline{\lambda})$.
Recalling $(\del\lambda\overline{\lambda})/(\lambda\overline{\lambda})\simeq \del \log (\lambda\overline{\lambda}) \simeq \del\varphi $,
one finally obtains the form of $\Bar{J}$,
that coincides with the one
obtained from the consistent gluing condition:
\begin{align}
  \Bar{J} &= -\varrho - {N-4\over2}\del\varphi\,.
\end{align}

\paragraph{Example: $b\omega_{i}$}

In the minimal BRST description $b\omega_{i}$ is BRST closed.
In fact, it is not difficult to check that it is in the
cohomology of $D$.
Now, just like $b$, $b\omega_{i}$ is also a representative
of a $\overline{D}$-cohomology.
But using the formulas
\begin{align}
\begin{split}
D\biggl( {(\overline{\lambda}\omega)\omega_{i} \over 2(\lambda\overline{\lambda})  }\biggr)
 &= b\omega_{i}
  + {b\lambda_{i}(\overline{\lambda}\omega) \over (\lambda\overline{\lambda})} +{\del b \overline{\lambda}_{i} \over (\lambda\overline{\lambda})} \,,
\\
D\biggl({  \lambda_{i} (\overline{\lambda}\omega)^{2} \over 4(\lambda\overline{\lambda})^{2} }\biggr)
 &= {b\lambda_{i}(\overline{\lambda}\omega) \over (\lambda\overline{\lambda})}
  + {\del b \overline{\lambda}_{i} \over (\lambda\overline{\lambda}) }\,,
\end{split}
 \end{align}
another representation of $b\omega_{i}$ (as an element of $(D+\deldol)$-cohomology)
can be obtained in which the ghosts are absent:
\begin{align}
\label{eq:bwinogh}
\begin{split}
  b\omega_{i} &\simeq
  -\deldol\biggl( { (\overline{\lambda}\omega)\omega_{i} \over 2(\lambda\overline{\lambda}) }
  - { \lambda_{i}(\overline{\lambda}\omega)^{2} \over 4(\lambda\overline{\lambda})^{2} } \biggr) \\
  &= { (\lambda r)(\overline{\lambda}\omega)\omega_{i} - (\lambda\overline{\lambda})(r  \omega)\omega_{i} \over 2(\lambda\overline{\lambda})^{2} }
  - { (\lambda r)(\overline{\lambda}\omega)^{2}\lambda_{i} - (\lambda\overline{\lambda})(r \omega)(\overline{\lambda}\omega)\lambda_{i} \over 2(\lambda\overline{\lambda})^{3} }\,.
\end{split}
\end{align}
Since $\deldol$ and $D$ commute,
the right hand side is necessarily gauge invariant (or $D$-closed).
Also, it is $\deldol$-closed being a $\deldol$ of
a gauge {\em non}-invariant ($D$-non-closed) operator,
but it cannot be written as a $\deldol$ of a gauge invariant operator.
Those implies that one can read-off
the corresponding element of the Dolbeault cohomology from~(\ref{eq:bwinogh}).
That is, with $(\omega,\lambda)$ being understood as constrained variables,
\begin{align}
\psi_i &= { (\lambda r)(\overline{\lambda}\omega)\omega_{i} - (\lambda\overline{\lambda})(r  \omega)\omega_{i} \over 2(\lambda\overline{\lambda})^{2} }
  - { (\lambda r)(\overline{\lambda}\omega)^{2}\lambda_{i} - (\lambda\overline{\lambda})(r \omega)(\overline{\lambda}\omega)\lambda_{i} \over 2(\lambda\overline{\lambda})^{3} }
\end{align}
is $\deldol$-closed
provided the quantum corrections are defined appropriately.
But it is not $\deldol$-exact and hence is in the Dolbeault cohomology.

In the \v{C}ech language.
the corresponding element can be found to be the $1$-cochain
\begin{align}
(\psi^{AB}_i)
&= {-2\lambda^{[A}\omega^{B]}\omega_{i} \over \lambda^{A}\lambda^{B}}
 +  {2\omega^{(A}\omega^{B)}\lambda_{i} \over \lambda^{A}\lambda^{B} } \,.
\end{align}
The argument for it being in the \v{C}ech cohomology is the same as the Dolbeault case.
It satisfies the cocycle condition on the triple overlaps $U_A\cap U_B \cap U_C$,
\begin{align}
(\psi^{AB}_i - \psi^{AC}_i + \psi^{BC}_i) = 0\,,
\end{align}
but it is not a coboundary of any gauge invariant operators,
and hence is in the \v{C}ech cohomology.

\section{Summary and discussion}
\label{sec:summary}

In this paper, we have studied the Hilbert space of the conformal field theories
with a simple quadratic constraint $\lambda^{i}\lambda^{i}$ ($i=1\sim N$)
using both curved $\beta\gamma$ (\v{C}ech/Dolbeault) and BRST frameworks.
Although there are slight mismatches
between the two descriptions due to the quantum ordering problem,
we found that their partition functions $\Tr[(-1)^{F}\cdots]$
agree for $N\ge4$ models.
Since our partition functions in both descriptions
are insensitive to quantum corrections,
the agreement of the partition functions can be explained by
classically relating the elements of the cohomologies
of the two formalisms.
We showed the classical equivalence of the two cohomologies
by embedding them into a combined bigraded cohomology.

Regarding the structure of the Hilbert space itself,
we found that the quantum BRST cohomology is non-vanishing only at
ghost numbers $0$ and $1$,
and that there is a one-to-one mapping between the two sectors.
In terms of the partition function, the mapping between ghost numbers $0$
and $1$ are summarized as the $\ast$-conjugation symmetry.
We explicitly constructed a non-degenerate inner product
that couples the two sectors.

In the BRST language, the lowest mass state in the ghost number $1$ cohomology
is accounted for by the ghost $b$ itself in the BRST operator $D=\int b(\lambda\lambda)$.
In Dolbeault language it corresponds to a $1$-form on the constrained surface,
and in \v{C}ech language it corresponds to a $1$-cocycle
defined only on the double overlaps of the coordinate charts.

\bigskip
There, however, are several points in the present work
that require further clarifications.
One of them is to understand
the discrepancy between the extrinsic (BRST)
and intrinsic (curved $\beta\gamma$) descriptions more precisely.

For the class of models we studied
(models on a cone over a base $B$ with the origin removed),
we encountered two sources for the discrepancy.
Firstly, for lower dimensional models ($N\le3$),
one finds operators that are globally defined
but nevertheless cannot be written as
a gauge gauge invariant polynomials in the extrinsic coordinates $(\omega,\lambda)$.
We presented an argument for the absence of such operators
when the base $B$ has dimensions greater than $1$ (i.e. $N\ge4$),
but it would be nice to understand the precise criterion.

At the quantum level,
second source for the discrepancy between
the BRST and curved $\beta\gamma$ descriptions
arises from the different normal ordering prescriptions used in the two.
A pair of the elements of the classical BRST cohomology can
drop out from the quantum cohomology by forming a BRST doublet,
$\Hat{g} = D\Hat{f}$.
In the curved $\beta\gamma$ framework, similar phenomenon occurs when
the quantum effect spoils
the gluing property of a classical cohomology $f$.
In this case, the failure of gluing is represented by
a higher cochain $g=\delcech f$ which is also in the classical cohomology.
Since the two frameworks use different normal ordering prescriptions,
there are discrepancies between the two phenomena.
It would be useful to study if this type of discrepancy can
be remedied, for example,
by appropriately bosonizing the BRST ghosts.

Another clarification that should be attempted is to
explore the one-loop path integral expression for
our partition functions.
When properly understood,
it should be useful for unconvering
the origin of the
field-antifield and $\ast$-conjugation symmetries.

\bigskip
Leaving the clarifications of those subtleties to a future work,
we list some directions for the extensions of the results
obtained in the present paper.

\bigskip
In an accompanying paper~\cite{PS}, we extend the result
to the more interesting case of pure spinors.
Despite the fact that the pure spinor constraint is infinitely reducible,
it will be argued that the structures above carry over almost literally.
The only difference is that the ghost numbers at which the cohomology become non-trivial
are $0$ and $3$, instead of $0$ and $1$.
The lowest mass state in the ghost number $3$ cohomology
carries weight $2$ and represents an important term in the reparameterization $b$-ghost.

Knowing that there can be no cohomologies with ghost numbers greater than $3$
is nice for the pure spinor multiloop amplitudes,
because it implies that one need not worry about the poles
coming from the fusion of many reparameterization $b$-ghosts.
The troublesome poles are necessarily carrying
ghost numbers greater than $3$ and,
modulo the subtleties coming from the divergences at the boundary of
moduli spaces,
they can be ignored without having have to use the regularization
introduced in~\cite{Berkovits:2006vi}.
It would be interesting to work out how it is actually realized,
and the present models might be useful to clarify
some aspects of this issue.

Finally, it should be possible to extend our results to the
curved $\beta\gamma$ systems with cubic or higher homogeneous constraints
(or intersections thereof).
For the case of single homogeneous constraint of order $L$,
the result is almost obvious.
The $\ast$-conjugation symmetry relates
the states with $q^{m}t^{n}g^{k}$ to those with $q^{m+n+{L(L-1)\over2}}t^{-n-L}g^{L-k-1}$
and it is not difficult to construct the inner product
that couples $H^{k}(D)$ and $H^{L-k-1}(D)$.
Note that all cohomologies $H^{k}(D)$'s with $0\le k\le L-1$
are non-empty
having $b\del b\cdots\del^{k-1}b$ as the lowest mass element.

\section*{Acknowledgments}

We would like to thank Nathan Berkovits and Nikita Nekrasov
for many useful discussions and encouragements.
Many of the results presented in the present work
owe them considerably.
The work of YA was supported by FAPESP grant 06/59970-5,
while that of EAA was supported by FAPESP grant 04/09584-6.

\section*{Appendix}
\appendix
\setcounter{equation}{0}
\def\thesection{\Alph{section}}
\renewcommand{\theequation}{A.\arabic{equation}}

\section{Curved $\beta\gamma$ system on the cone $\lambda^{i}\lambda^{i}=0$}
\label{app:geomcone}

In this appendix, we collect some useful formulas
for the study of the curved $\beta\gamma$ system on
the $N-1$ (complex) dimensional cone
\begin{align}
  X &= \{ \lambda^{i} \;|\; \lambda^{i}\gamma_{ij}\lambda^{j} = 0\,,\quad \lambda\ne0 \}  \subset \mathbb{C}^{N}\,,\quad(i,j=1\sim N)\,.
\end{align}
Here, $\gamma_{ij}$ is a constant symmetric ``metric''.
Below, we diagonalize $\gamma_{ij}$ and do not distinguish upper and lower
indices.
Also, we always assume that the origin $\lambda=0$ is removed
so $X$ is a $\mathbb{C}^{\ast}$-bundle over a base $B$.

\subsection{Geometry of the cone $\lambda^{i}\lambda^{i}=0$}

\subsubsection{An open covering}

Let us denote
\begin{align}
  \lambda^{I} &= \lambda^{i}+\mathi\lambda^{i+1}\,,\quad\lambda^{\Tilde{I}}=\lambda^{i}-\mathi\lambda^{i+1}\,.
\end{align}
Here, $i$ runs over $1\sim N$ and is defined modulo $N$.
We also use the index $A$ to denote both $I$ and $\Tilde{I}$
and use notations
\begin{align}
  \lambda^{A} = (\lambda^{I}, \lambda^{\Tilde{I}})\,,\quad
  \lambda_{A}={1\over2}(\lambda^{\Tilde{I}},\lambda^{I})\,,\quad
  \sum \lambda^{A}\lambda_{A} = \sum \lambda^{I}\lambda^{\Tilde{I}} = \lambda^{i}\lambda^{i}\,.
\end{align}
The cone $X$ can be covered by $2N$ patches $\{U_A\}_{A=1\sim 2N}$,
where on a patch at least one of $\lambda^{A}$ is non-vanishing:
\begin{align}
U_{A} = \{\lambda \;|\; \lambda^{A}\ne0\}
\quad\leftrightarrow\quad
  U_{I} = \{ \lambda \;|\; \lambda^{I} \ne 0\} \;\;\text{or}\;\;
  \Tilde{U}_{\Tilde{I}} = \{\lambda\;|\; \lambda^{\Tilde{I}}\ne0 \}\,.
\end{align}
On a patch, $\lambda$ can be parameterized using $N-1$ independent variables $(g,u^{a})$,
where $g$ parameterizes the overall scale of $\lambda$,
and $u^{a}$'s are $N-2$ ``angular'' variables.
For example, on $U_1$ and $\Tilde{U}_{\Tilde{1}}$,
the local coordinates are $(g_{(1)},u_{(1)}^{a})_{a=3\sim N}$
and $(\Tilde{g}_{(1)},\Tilde{u}_{(\Tilde{1})}^{a})_{a=3\sim N}$ respectively,
and $\lambda$ is parameterized as (omitting the subscript $(1)$ and $(\Tilde{1})$ for simplicity)
\begin{align}
\begin{split}
  U_1\colon (\lambda^{1}+\mathi\lambda^{2},\lambda^{1}-\mathi\lambda^{2},\lambda^{a})
  &= (g,g(u\cdot u),\mathi g u^a)\,,\\
  \Tilde{U}_1\colon (\lambda^{1}+\mathi\lambda^{2},\lambda^{1}-\mathi\lambda^{2},\lambda^{a})
  &= (\Tilde{g}(\Tilde{u}\cdot\Tilde{u}),\Tilde{g},\mathi \Tilde{g} \Tilde{u}^a)\,.
\end{split}
\end{align}
Variables $(g,u^{a})$ on other patches are defined in a similar manner.

\subsubsection{Coordinate transformation}

The transformations among the coordinates above are readily computed.
We here give
the transition functions on $U_1\cap \Tilde{U}_{\Tilde{1}}$,
$U_1\cap U_{2}$ and $U_{\Tilde{1}}\cap U_{2}$.

On the overlap $U_1\cap \Tilde{U}_1$,
both $\lambda^{1}+\mathi\lambda^{2}=g=\Tilde{g}(\Tilde{u}\cdot\Tilde{u})$
and $\lambda^{1}-\mathi \lambda^{2}=\Tilde{g} = g(u\cdot u)$
are non-vanishing.
Hence, $(u\cdot u)$ and $(\Tilde{u}\cdot \Tilde{u})$ are
also non-vanishing and the two coordinates are related by
\begin{align}
  (g,u^a) &= (\Tilde{g}(\Tilde{u}\cdot \Tilde{u}),\Tilde{u}^a(\Tilde{u}\cdot\Tilde{u})^{-1})\,,
 \quad
  (\Tilde{g},\Tilde{u}^a)= (g(u\cdot{u}),{u}^a({u}\cdot{u})^{-1})\,.
\end{align}

To describe the transformation on the
overlap between $U_{1}$ and $U_2$,
let us temporarily denote
\begin{align}
  (G,U^{a}) &= (g_{(2)},u_{(2)}^{a})\,,\quad (a=4\sim (N+1)=1,4\sim N)\,.
\end{align}
On the overlap $U_1\cap U_2$,
$g$ and $G$ as well as
$(1 - 2\mathi u_3 -u\cdot u)$
and $(1 + 2U_1 +U\cdot U)$
are non-vanishing and $(g,u^a)$ and $(G,U^a)$ are related as
\begin{align}
\begin{split}
  g &= {\mathi\over2}G(1+2U_1+U\cdot U)\,,\quad
  u_3 = {\mathi(1-U\cdot U )\over 1 + 2U_1 +U\cdot U}\,,\quad
  u_a = { 2\mathi U_a \over 1+2U_1+U\cdot U}\,,
\\
G &= -{\mathi\over2}g(1-2\mathi u_3-u\cdot u)\,,\quad
U_1 = {1+u\cdot u \over 1 - 2\mathi u_3 -u\cdot u}\,,\quad
U_a = { 2\mathi u_a \over 1-2\mathi u_3-u\cdot u}\,.
\end{split}
\end{align}

Similarly, the relation between $(\Tilde{g},\Tilde{u})$ and $(G,U)$
on the overlap $\Tilde{U}_{\Tilde{1}}\cap U_2$ are given by
\begin{align}
\begin{split}
\Tilde{g} &= -{\mathi\over2}G(1-2U_1+U\cdot U)\,,\quad
\Tilde{u}_{3} = { -\mathi(1-U\cdot U)\over 1-2U_1+U\cdot U } \,,\quad
\Tilde{u}_{a}= { 2\mathi U_a\over 1-2U_1+U\cdot U }\,,
\\
G &= {\mathi\over2}\Tilde{g}(1+2\mathi \Tilde{u}_3 -\Tilde{u}\cdot \Tilde{u})\,,\quad
U_1 = { -1-\Tilde{u}\cdot\Tilde{u}\over 1+2\mathi\Tilde{u}_3-\Tilde{u}\cdot\Tilde{u}}\,,\quad
U_a = { -2\mathi\Tilde{u}\cdot\Tilde{u}\over 1+2\mathi\Tilde{u}_3-\Tilde{u}\cdot\Tilde{u}}\,.
\end{split}
\end{align}
One can easily check the consistency of the transformations
on the triple overlap $U_{1}\cap U_{\Tilde{1}}\cap U_{2}$.

\subsubsection{Partition of unity}

By introducing the non-minimal variables $\overline{\lambda}_{i}$
(complex conjugates to $\lambda^{i}$),
a partition of unity on $X$ can be constructed explicitly as
\begin{align}
  \rho_{A} = {\lambda^{A}\overline{\lambda}_{A} \over \lambda\overline{\lambda}}\,,\quad
(\lambda\overline{\lambda}=\lambda^{i}\overline{\lambda}_{i}=\sum_{A}\lambda^{A}\overline{\lambda}_{A}=\sum_{A}g_{(A)}\overline{g}_{(A)})\,.
\end{align}
Clearly, $\{ \rho_{A} \}$ is subordinate to the covering $\{U_{A}\}$,
that is, $\rho_{A}=0$ outside the patch $U_{A}$.
The derivative of $\rho_{A}$ is
\begin{align}
  \targetdbar\rho_{A} = {(\lambda\overline{\lambda})r_{A}\lambda^{A} - (\lambda r)\overline{\lambda}_{A}\lambda^{A}\over (\lambda\overline{\lambda})^{2}} \,.
\end{align}
A \v{C}ech $n$-cochain $(f^{A_0A_1\cdots A_{n}})$ and the corresponding $n$-form in Dolbeault language
$\Bar{f}$ are related as
\begin{align}
  \Bar{f} &= {1\over (n+1)!}\sum f^{A_0A_1\cdots A_{n}} \rho_{A_0}\targetd \rho_{A_1}\cdots \targetd \rho_{A_n}\,.
\end{align}

\subsection{$\beta\gamma$ system on the cone $\lambda^{i}\lambda^{i}=0$}

\subsubsection{Free curved $\beta\gamma$ system on a patch}

On a patch $U_{A}$,
the conjugates to $(g,u^{a})$ are denoted as $(h,v_a)$ and they satisfy
the free field operator product expansions
\begin{align}
  h(z)g(w) &= {-1\over z-w}\,,\quad
  v_{a}(z)u^{b}(w) = {-\delta_{a}{}^{b} \over z-w}\,.
\end{align}
Since $g$ is non-vanishing, one can instead use $\varphi=\log g$ and its conjugate $\varrho$
satisfying
\begin{align}
  \varrho(z)\varphi(w) &= {-1\over z-w}\,.
\end{align}

\subsubsection{Transformation of momenta}

On an overlap $U_{A}\cap U_{B}$, the momenta on $U_{A}$ and those on $U_{B}$
are related as
\begin{align}
\label{eq:app:vtransf}
  \vec{v}_{(B)} &= \NO{ \vec{v}_{(A)} (\tau_{AB}) }+ (\phi_{AB})\del\vec{u}_{(A)} \,,
\end{align}
where we denoted $\vec{u}_{(A)}= (\varphi_{(A)}, u_{(A)})$
and $\vec{v}_{(A)}= (\varrho_{(A)}, v_{(A)})$
for simplicity.
$(\tau_{AB})$ is the Jacobian $(\del \vec{u}_{A} / \del \vec{u}_{B})$,
and the matrix $(\phi_{AB})$ is defined so that $\vec{v}_{(B)}$'s do not have
singular operator products among themselves.

On the overlap $U_{1}\cap \Tilde{U}_{\Tilde{1}}$,
the momenta are related as
\begin{align}
\begin{split}
  \Tilde{\varrho} &= \varrho - {(N-4)\over2}\del\log (u\cdot u)\,,\\
  \Tilde{v}_{a} &= 2\varrho u_{a} + (u\cdot u)v_{a} - 2(u \cdot v)u_{a} + 4\del u_a -(N-4)(\del \varphi)u_{a}
 \,.
\end{split}
\end{align}
Since $\Tilde{v}_{a}$ generates a translation on $\Tilde{U}_{\Tilde{1}}$,
it should agree with the corresponding rotation generator $N^{-}_{a}$
in the coordinate $U_{1}$.
This indeed is the case (see below).

On the overlap $U_1 \cap U_2$, the momenta in $U_{2}$ which we denote
$(R,V_{1},V_{a'})_{a'=4\sim N}$ are
\begin{align}
\begin{split}
R &= \varrho - {(N-4)\over4}\del\log(1-2\mathi u_{3} - u_{3}^{2} + u_{c'}u_{c'}) \,,
\\
V_{1} &= (1-\mathi u_3)(\varrho - v_{c'}u_{c'})
  - {\mathi\over2}(1-2\mathi u_3 - u_{3}^2 +u_{c'}u_{c'})v_3  \\
&\qquad  -2\mathi\del u_{3} - {(N-4)\over2}(1-\mathi u_{3})\del\varphi \\
&= -N + {\mathi\over2}N_{3}^{+} + {\mathi\over2}N^{-}_{3} \,,
\\
V_{a'} &=
 \mathi \varrho u_{a'} + (v_3 - \mathi v_3u_3 - \mathi v_{c'}u_{c'})u_{a'}
  - {\mathi\over2}(1-2\mathi u_{3} - u_3^2 - u_{c'}v_{c'})v_{a'}  \\
&\qquad
   +2\mathi \del u_{a'}
   -{(N-4)\mathi \over2}u_{a'}\del\varphi \\
&= {\mathi\over2}N^{+}_{a'} - N_{3a'} - {\mathi\over2}N^{-}_{a'} \,.
\end{split}
\end{align}
Again, $V_{1,a'}$ corresponds to certain
linear combinations of the rotation currents.

\bigskip
The quantum correction part
$(\phi_{AB})\del \vec{u}_{A}$ in~(\ref{eq:app:vtransf})
cannot be defined consistently to satisfy the cocycle condition
$(\phi_{AC})(\phi_{BC})(\phi_{AB})=1$,
unless a closed $2$-form valued $2$-cocycle
\begin{align}
(\psi_{ABC}) = \tr(\tau_{AB}\wedge \targetd \tau_{BC} \wedge \targetd \tau_{CA})
\end{align}
represents a trivial class in the \v{C}ech cohomology.
On the triple overlap $U_1\cap\Tilde{U}_{\Tilde{1}}\cap U_2$,
$\psi$ is given by
\begin{align}
\begin{split}
\label{eq:app:anomtwo112}
  (\psi_{1\Tilde{1}2}) &= \tr(\tau_{1\Tilde{1}}\wedge \targetd\tau_{\Tilde{1}2}\wedge\targetd \tau_{21}) \\
  &= \sum_{a'=4}^{N}{4\mathi(N-4) u_{a'} \targetd u_{3} \wedge \targetd u_{a'}
   \over (u\cdot u)(1-2\mathi u_3 + u\cdot u)} \\
  &=  (N-4)\targetd\log(u\cdot u) \wedge \targetd\log(1-2\mathi u_3 + u\cdot u)  \,.
 \end{split}
\end{align}
This expression of $\psi$ tells us two things.
First, note that the right hand side only includes the coordinates
of the base $B$.
This is a general feature of the models with a $\mathbb{C}^{\ast}$-fiber
and $\psi_{1\Tilde{1}2}$ coincides with the obstruction for the
model on the base $B$.
On $B$, there is no way to rewrite~(\ref{eq:app:anomtwo112})
as a coboundary of a $2$-cochain
holomorphic in $U_{A}\cap U_{B}$ (restricted to $B$),
so the curved $\beta\gamma$ system with target space $B$ is anomalous,
i.e. the momenta cannot be glued consistently.

At the same time, we find from~(\ref{eq:app:anomtwo112})
that $\psi$ is in fact trivial on $X$,
as it is a coboundary of $2$-cochains
holomorphic in $U_{1}\cap U_{\Tilde{1}}$, $U_{1}\cap U_{2}$ and $U_{\Tilde{1}}\cap U_{2}$:
\begin{align}
\begin{split}
\psi_{1\Tilde{1}2}
&\propto \delcech(\targetd\varphi\wedge\targetd\Tilde{\varphi},\, \targetd\varphi\wedge \targetd\Phi,\,
       \targetd\Tilde{\varphi}\wedge\targetd\Phi) \\
&=  \targetd \varphi \wedge \targetd\log(u\cdot u)
  - \targetd\varphi \wedge \targetd\log(1-2\mathi u_3 + u\cdot u)  \\
 &\qquad
  +\targetd(\varphi + \log(u\cdot u))\wedge
    \targetd(\varphi + \log(1-2\mathi u_3 + u\cdot u) ) \\
 &=  \targetd\log(u\cdot u) \wedge \targetd\log(1-2\mathi u_3 + u\cdot u)  \,.
\end{split}
\end{align}
That is, the obstruction $(\psi_{ABC})$ represents a trivial class
$\delcech(\targetd\varphi_A\wedge\targetd\varphi_B)$ in the \v{C}ech cohomology,
so the momenta on $X$ (unlike those restricted on $B$)
can be glued consistently.

\subsubsection{Symmetry currents}

\label{app:JN}

The cone $X$ is invariant under the rescaling and rotations of $\lambda$.
In a given patch, the corresponding currents take the following forms:
\begin{align}
\begin{split}
  J &= -\varrho - {n-4\over 2}\del\varphi\,, \\
  N &= (v\cdot u)  - J'\,,\quad  N_{ab} = -v_{a}u_{b} + v_{b}u_{a} \,,\\
  N^{+}_{a} &= -v_{a}\,,\quad N^{-}_{a} = 2(v\cdot u)u^{a} -(u\cdot u)v^{a} - 2J' u^{a} - 4\del u^{a}\,.
\end{split}
\end{align}
Here, $J'=\varrho-{n-4\over2}\del\varphi$ is defined so that $J(z)J'(w)$ have no poles,
and we temporarily denoted the number of $\lambda$ components by $n$,
to avoid the confusion with the operator $N$ that generates $U(1)\subset SO(n)$.

$(J,N)$ form the $U(1)_{t}\times SO(n)$ current algebra
with levels $(4-n,-2)$:
\begin{align}
\begin{split}
  J(z) J(w) &= {4-n \over (z-w)^{2}} \,, \\
  N(z)N(w) &= {-2\over (z-w)^{2}}\,,\quad
  N(z)N^{\pm}_{a}(w) = {\pm N^{\pm}_{a}(w)\over z-w} \,,\\
  N^{+}_{a}(z)N^{-}_{b}(w) &= { -4\delta_{ab}\over (z-w)^{2}}  + { 2N_{ab}(w) + 2\delta_{ab}N(w) \over z-w}\,, \\
  N^{+}_{a}(z)N_{bc}(w) &= {2\delta_{a[b}N^{+}_{c]}(w) \over z-w}\,,\quad
  N^{-}_{a}(z)N_{bc}(w) = {-2\delta_{a[b}N^{-}_{c]}(w) \over z-w} \,, \\
  N_{ab}(z)N_{cd}(w) &= { -2(\delta_{ad}\delta_{bc}-\delta_{ac}\delta_{bd}) \over (z-w)^{2}} + {2\delta_{a[c}N_{d]b}(w)-2\delta_{b[c}N_{d]a}(w) \over z-w}\,, \\
  (\text{others}) &= \text{regular}\,.
\end{split}
\end{align}
Note that the rescaling by $J$ commutes with the rotations by $N$.

\subsubsection{Energy-momentum tensor}
\label{app:T}

Finally, using the coordinate above, one can construct the nowhere vanishing
holomorphic top form $\Omega$ on $X$.
Choosing the orientation of the coordinates consistently,
it takes the form
\begin{align}
  \Omega &= \mathe^{(N-2)\varphi}\targetd\varphi\wedge\targetd u^{3}\wedge\cdots \wedge \targetd u^{N}\,,
\end{align}
in all coordinate patches.
Definition of $\Omega$ is purely geometric
and it is straightforward to check that it transforms covariantly on the
overlaps.
Hence, $X$ is a (non-compact) Calabi-Yau space
and one can define a globally defined conformal field theory
for which the energy-momentum tensor is given by gluing
\begin{align}
\begin{split}
  T &= T_{\text{naive}}- {1\over2}\del^{2}\log(\mathe^{(N-2)\varphi}) \\
   &= -\varrho\del\varphi - v_a\del u^a - {(N-2)\over2}\del^{2}\varphi\,.
\end{split}
\end{align}
Note that the background charge for $\varphi$ obtained here
is consistent with the $t$-charge anomaly
\begin{align}
  J(z)T(w) &= {2-N \over (z-w)^{3}} + {J(w)\over (z-w)^{2}}\,.
\end{align}


\end{document}